\documentclass[11pt]{article}

\usepackage{amssymb}
\usepackage{amsmath}
\usepackage{graphicx}
\renewcommand{\vec}[1]{{\bf #1}}    
\def\beq{\begin{eqnarray}}       
\def\eeq{\end{eqnarray}}        

\def\Box{\square}           
\def\diag{\,\mbox{diag}\,}           

\newcommand\pubblock{\rightline{\begin{tabular}{l} \pubnumber\\
     \pubdate\\ \hepnumber \end{tabular}}}
\newcommand\pubnumber{UB-ECM-PF-08/14}
\newcommand\pubdate{ }
\newcommand\hepnumber{}

\def\al{\alpha}
\def\be{\beta}

\def\ga{\gamma}
\def\de{\delta}
\def\ep{\epsilon}

\def\La{\Lambda}
\def\la{\lambda}

\def\si{\sigma}
\def\om{\omega}

\def\Ga{\Gamma}
\def\De{\Delta}
\def\La{\Lambda}

\def\na{\nabla}
\def\pa{\partial}

\newcommand{\rL}{\rho_{\Lambda}}
\newcommand{\CC}{\Lambda}
\newcommand{\rLO}{\rho^0_{\Lambda}}
\newcommand{\rLV}{\rL^{\rm vac}}
\newcommand{\rLI}{\rL^{\rm ind}}
\newcommand{\EPR}{V_{eff}}

\begin{document}
\pubblock

\begin{center}

{\large\sc Can the cosmological ``constant'' run? - It may run}
\vskip 8mm

\textbf{Ilya L. Shapiro}$^{\,a}$\footnote{%
E-mail: shapiro@fisica.ufjf.br}
$\,,\,\,$
\textbf{Joan Sol\`{a}}
$^{\,b\,,c\,,}$\footnote{E-mail: sola@ifae.es.}
\vskip 12mm
$^{\,a}$
Departamento de F\'{\i}sica - Instituto de Ciencias Exatas
\\
Universidade Federal de Juiz de Fora, 36036-330, MG, Brazil \vskip
2mm $^{\,b}\,$ HEP Group, Dept. E.C.M.  Universitat de Barcelona,
\\
\,Diagonal 647, Barcelona, Catalonia, Spain \vskip 2mm $^{\,c}\,$
Institut de Ci\`encies del Cosmos, Universitat de Barcelona
\\[0pt]
\end{center}
\vspace{0.0 cm}

\begin{center}
{ABSTRACT}
\end{center}

\begin{quotation}
\noindent Using standard quantum field theory, we discuss several
theoretical aspects of the possible running of the cosmological
constant (CC) term in Einstein's equations. The basic motivation for
the present work is to emphasize that this possibility should also
be taken into account when considering dynamical models for the dark
energy (DE), which are nowadays mainly focused on identifying the DE
with the energy density associated to one or more \textit{ad hoc}
scalar fields. At the same time, we address some recent criticisms
that have been published (or privately communicated to us)
attempting to cast doubts on the fundamental possibility of such
running. In this work, we argue that while there is no comprehensive
proof of the CC running, there is no rigorous proof of the
non-running either. In particular, some purported ``non-running
theorem'' recently adduced in the literature is, in our opinion,
completely insubstantial and formally incorrect. The way to the CC
running is, therefore, still open and we take here the opportunity
to present a pedagogical review of the present state of the art in
this field, including a brief historical account.\vskip 1mm

\end{quotation}

\vskip 8mm

\section{\large\bf Introduction}

\qquad The relevance of the cosmological constant (CC) problem (in
its various aspects\,\cite{weinberg89}) has triggered a vast
interest in identifying the possible quantum effects on the vacuum
energy density and their potential implications in cosmology. Many
of the proposed approaches to the CC problem are based on
considering that the vacuum energy is related, or even superseded,
by the specific properties of some ad hoc scalar field (or
collection of them)\,\cite{DEquint}. This point of view is what has
finally led to the general notion of dark energy (DE) in its
manifold possible forms\,\cite{DE}. However, in this work, we will
essentially concentrate on the cosmological constant term in
Einstein's equations and, more generally, on the quantum field
theoretical aspects of the renormalizable vacuum action that ensues
after appropriately extending the original Einstein-Hilbert action.
This, more fundamental, QFT point of view naturally leads to
consider the renormalization group (RG) properties of the parameters
of the renormalizable vacuum action, in particular of the CC term,
$\CC$, in it. One expects that the physical CC resulting from this
RG approach can be viewed as a running entity that could perhaps
explain, in truly fundamental terms, the suspected dynamical nature
of the DE, which has long been invoked in order to explain other
intriguing aspects of the CC conundrum, such as the cosmic
coincidence problem\,\cite{DEquint,DE}.

There are already many papers on the subject under consideration,
and one can distinguish different approaches. First of all, there is
the great CC problem or ``old CC problem'' (approach {\it A}), i.e.
the formidable task of trying to explain the relatively small (for
Particle Physics standards) measured value of the CC\,\cite{CCdata},
roughly $\rL^{\rm exp}\sim 10^{-47}\,GeV^4$, after the many phase
transitions that our Universe has undergone since the very early
times\,\cite{weinberg89}. Second, we have the approach to the CC
problem relying on the renormalization group method (approach {\it
B}), namely by exploring the possibility that a moderate running of
the CC could leave a physical dynamical imprint that can eventually
be observed\,\cite{Irgac}. Notice that this RG viewpoint could also
have some bearing on the aforementioned cosmic coincidence problem,
see\,\cite{JCAPAna1}. Let us remark, however, that approach {\it B}
has no obvious impact on solving the main problem addressed in
approach {\it A}.

On the other hand, there is a dramatic distinction between the
``RG-cosmologies'' based on the renormalization group properties of
the unknown fundamental physics such as quantum gravity or string
theory (sub-approach {\bf I}) and the  alternative renormalization
group approach based on the familiar methods and results of the
Standard Model (SM) of Particle Physics (sub-approach {\bf II}) and
its favorite extensions (such as e.g. the Minimal Supersymmetric
Standard Model\,\cite{MSSM}). While the implications of the RG
method in sub-approach {\bf I} are difficult to test in practice, in
the case {\bf II} the RG technique relies on the well-tested facts
of the SM and/or of its extensions which, by the way, will be soon
put to the test in the upcoming generation of $TeV$-class colliders,
headed by the imminent startup of the Large Hadron Collider at CERN.

It is pretty obvious that the philosophy of the various combined
approaches, e.g., {\it AI} and {\it BII}, is quite different. In the
first case, we meet an attempt to solve the old CC problem using the
assumed, but yet unknown, fundamental theory of everything (TOE).
The works in this direction deserve a great respect, even though the
creation of a successful and falsifiable fundamental TOE looks a
rather remote perspective for the time being, exactly as the
solution of the great CC problem \cite{weinberg89}. In the second
case ({\it BII}), the purpose is more modest and, perhaps for this
reason, more viable; namely, it consists to explore whether the
observable CC (whose value is not attempted to be accounted for) can
be mildly variable due to the quantum effects generated from the
well established quantum field theory (QFT) of Particle Physics,
such as the SM and extensions thereof, but in the presence of an
external (and dynamical) metric background. The exploration of this
possibility, even if looks less ambitious, it represents
nevertheless a necessary step on the way to find a fundamental
physical explanation for the vacuum energy problems. In this
article, we are going to discuss, in a bit more detail than ever
before, the quantum field theory aspects of a possible
renormalization group running of the vacuum energy.

The paper is organized as follows. In section 2 we present, in a
pedagogical way, the preliminaries of the CC running, that is to
say, we explain what is the cosmological constant and what means the
renormalization group running. In section 3, we present a brief (and
probably incomplete) history of the RG for the vacuum action
parameters, and especially for the cosmological constant. We shall
provide classification (e.g. {\it AI} or {\it BII}, etc) to each of
the approaches.

Despite that many of the considerations presented here are rather
general, from section 4 on we will concentrate on the approach {\it
BII} and discuss the possibility of having moderate CC running owing
to the quantum effects of massive matter fields (e.g., SM
constituents). This approach has been first started in
\cite{cosm,nova} and further developed in recent years
\cite{babic,GHScosm,CCfit,IRGA03,Babic2,Gruni,JSHSPL05,FSS1}. Our
aim here is to clarify its theoretical background in a consistent
form. It is worthwhile to say, from the very beginning, that we do
not have a comprehensive proof that the scale dependence (running)
of the CC really takes place, but at the same time we emphasize that
there is no correct proof of the opposite. Therefore, from our point
of view, the relevance of the subject makes perfectly reasonable any
sound investigation in this area whether performed within
phenomenological or theoretical terms.

Notwithstanding, it is also very important to distinguish these
sound investigations from conceptually wrong results in the
literature (of which we will comment in detail later on) and thus
maintain an unbiased approach to the possibility of CC running. For
example, in section 4, we explain the important relationship of the
effective potential and its RG invariance with the cosmological
constant. It suffices to say here that the ``non-running'' claim
adduced in a recent publication\,\cite{AG} on the sole basis of the
RG-invariance of the effective potential, is incorrect and entirely
misleading from the conceptual and technical point of view. We
explain the mistakes on which that work is based, specially due to
the misuse of the RG approach in QFT. Moreover, the technical
simplicity of this section enables us to provide a kind of general
introduction to the subject, which will hopefully be interesting to
the readers.

In section 5 we address other (more serious) arguments against the
variable CC, including the ones presented recently by A. Vilenkin
\cite{Vilenkin} and A.D. Dolgov \cite{Dolgov}. We hope to convince
the reader that, despite these and other arguments (including our
own) challenging the possibility of a variable CC, there is still an
open window that can accommodate the restricted set of quantum
corrections which can provide the physical support for such a
running. Finally, in the last section we draw our conclusions.

\section{Background notions}

\qquad Before starting to discuss the possibility and meaning of
having CC running, and ultimately of having ``RG Cosmology'', it is
worthwhile to remember what is the definition of the CC, what is the
energy of vacuum and what is the renormalization group running in
QFT.

\subsection{What is  $\CC$}

\qquad The modern theory of gravitation starts from a classical
action that goes beyond the conventional Einstein-Hilbert (EH)
action. It includes a more complete structure for the vacuum part,
together with the matter:
\beq S_{total} = -\frac{1}{16\pi G}\,\int d^4 x\,\sqrt{-g} \,\left(
R + 2\La \right) + S_{HD} + S_{matter}\,. \label{total} \eeq
Here, the first term on the \textit{r.h.s.} is the EH action with
the cosmological constant $\La$; the second term ($S_{HD}$) includes
the higher derivatives that are necessary for the consistency
properties (such as renormalizability) of the quantum theory in
curved space (see, e.g., \cite{birdav}, \cite{book} or \cite{PoImpo}
for an introduction) and the last term represents the action of
matter, which is responsible for the energy-momentum tensor
$T_{\mu\nu}^M=-2(-g)^{-1/2}\delta S_{matter}/\delta g^{\mu\nu}$. It
is also customary to include the contribution of the higher
derivative term $S_{HD}$ into the energy-momentum tensor; hence the
full energy-momentum tensor associated to the action (\ref{total})
reads $T_{\mu\nu}=
-2(-g)^{-1/2}\delta\left(S_{HD}+S_{matter}\right)/\delta
g^{\mu\nu}$. Thus, in the absence of matter, we have the vacuum
value $\,T_{\mu\nu}^{vac} = - 2(-g)^{-1/2}\delta S_{HD}/\delta
g^{\mu\nu}$. This custom gave origin, for instance, to the
expression ``trace anomaly'', given by $\langle
T_{\mu}^{\mu}\rangle$. This is perhaps an abuse of language, but we
shall follow the tradition and use the same wording.

Since our main purpose here is the description of the low-energy
gravitational physics, we shall at the moment disregard the higher
derivative piece $S_{HD}$ until we start to discuss the quantum
effects in curved space-time (see section 5). Therefore, the low
energy gravitational physics is well described by just the Einstein
dynamical equations for the metric:
\beq R_\mu^\nu - \frac12\,R\,\de_\mu^\nu \,=\,8\pi G\,T_\mu^\nu +
\La\,\de_\mu^\nu\,, \label{Einstein} \eeq
where $T_{\mu\nu}$ here is taken to be $T^M_{\mu\nu}$. Let us now
consider a cosmological system of matter representing our Universe.
According to the Cosmological Principle, on average it is described
by an isotropic and homogeneous fluid, with energy-momentum tensor
${T}^\nu_\mu= -p\,\delta^\nu_\mu+(\rho+p)U^{\nu}U_{\mu}\,.$
\label{Tmunuideal}
Therefore, in the locally co-moving frame, where the $4$-velocity of
the fluid is $U^{\mu}=(1,0,0,0)$, it takes the simplest form
\beq T^\nu_\mu=\diag \left(\rho,\,-p,\,-p,\,-p\right)\,, \label{EMT}
\eeq
where $\rho$ and $p$ are the proper density and pressure of matter.
It is easy to see that the $\La$ term on the \textit{r.h.s.} of
(\ref{Einstein}) can be written $\La\,\de_\mu^\nu =
8\,\pi\,G\,{\left(T_{\Lambda}\right)}^\nu_\mu$, where the new
energy-momentum tensor ${\left(T_{\Lambda}\right)}^\nu_\mu$ takes
the form (\ref{EMT}), with the ``vacuum energy density''
$\rho_\Lambda$ and ``vacuum pressure'' $p_\Lambda$ related as
\beq \rho_\Lambda = \frac{\Lambda}{8\pi G} = - p_\Lambda\,.
\label{CCd2} \eeq
In other words, we have ${\left(T_{\Lambda}\right)}^\nu_\mu=
\rL\,\delta^\nu_\mu$, which justifies $\rL$ being called the
``vacuum energy'', or at least a contribution to it.  So we shall
also use this terminology in what follows.

In order to better understand the role of the curved space-time
background for the relevance of the CC term, let us consider the
following example. Suppose we take another parametrization (Weyl
scaling) for the
metric %
\beq g_{\mu\nu} = \frac{\chi^2}{M^2}\, {\bar g}_{\mu\nu}\,,
\label{conf_rep} \eeq
where ${\bar g}_{\mu\nu}$ is some fiducial metric with fixed
determinant. For instance, it can be a flat metric $\,{\bar
g}_{\mu\nu}=\eta_{\mu\nu}$. Furthermore, $\chi=\chi(x^\mu)$ is a
scalar field which can be identified as a conformal factor of the
metric, $M$ is some mass (perhaps near the Planck mass, $M_P$). It
is easy to see that the CC term looks rather different in these new
variables
\beq S_\Lambda \,=\, -\, \int d^4x \sqrt{- g}\, \frac{\Lambda}{8\pi
G} \,=\, -\, \int d^4x \sqrt{-{\bar g}}\,f \chi^4\,, \qquad
\mbox{where} \qquad f = \frac{\Lambda}{8\pi G\,M^4}\,. \label{CCd1}
\eeq
The last action is nothing else than the usual potential term for
the scalar interaction in $\chi^4$ theory. Equivalently: the inverse
image of the $\chi^4$  interaction by the Weyl transformation is an
effective CC term in the Einstein frame. One can note that, under
the same change of variables, the Einstein-Hilbert term transforms
into the action of the scalar field $\chi$ with the negative kinetic
term and non-minimal conformal coupling
\cite{deser,sola8990,conf,Shocom}. Furthermore, the massive term in
the spinor Lagrangian becomes a Yukawa-type interaction between the
fermion and the scalar degree of freedom of the metric $\chi$. If we
are interested, e.g., in the renormalization of the gravitational
action due to the quantum effects of the spinor field, the procedure
is completely similar to the well known Nambu-Jona-Lasinio (NJL)
model. In this model, one does not quantize the scalar field,
exactly as we do not intend to quantize the metric in the approach
{\it BII}. Let us notice in passing that, within the Minimal
Subtraction (MS) scheme of renormalization, the RG running of the
cosmological constant in the theory of the spinor field is
mathematically equivalent to the one of the parameter $\,f\,$ in the
NJL model \cite{NJL}. A similar equivalence can be easily
established for the quantum effects of a free massive scalar field
$\phi$ -- in this case the quantum field interacts with $\,\chi\,$
via the $\phi^2\chi^2$-term.


 \subsection{Cosmologies based on a 
``running'' $\CC$}

Next, let us concentrate on the CC term and start by making an
important remark, which is at the heart of the message that this
paper wishes to convey to the reader: the CC is a nontrivial
quantity in field theory \textit{only} when there is a nontrivial
metric. Let us first consider the classical field equations. The
Bianchi identity fulfilled by the Einstein's tensor on the
\textit{l.h.s.} of (\ref{Einstein}), tells us that
\begin{equation}
\label{CCconstant} \nabla^{\mu}\Big(R_\mu^\nu -
\frac12\,R\,\de_\mu^\nu \Big)=0\,, \qquad \mbox{which implies}
\qquad \nabla^{\mu}\left(8\pi G\,T_\mu^\nu +
\La\,\de_\mu^\nu\right)=0\,.
\end{equation}
Thus, if $S_{HD}$ is negligible, $G$ is constant and matter is
covariantly conserved ($\nabla_{\nu} T^\nu_\mu =0$), this requires
$\Lambda$ to be a constant (i.e. a general coordinate-invariant
expression). Let us also notice that, if matter is non-covariantly
conserved, equation (\ref{CCconstant}) can still be satisfied
through a special local conservation law that enforces transfer of
energy between matter and the vacuum energy density (\ref{CCd2}).
Indeed, using the FLRW metric, characterized by the scale factor
$a=a(t)$, it is easy to see that the following
 relation holds \,\cite{Gruni}:
\begin{equation}\label{BianchiGeneral}
\frac{d}{dt}\,\big[G(\rho_\Lambda+\rho)\big]+3\,G\,H\,(\rho+p)=0\,,
\end{equation}
where $H=\dot{a}/a$ is the expansion rate or Hubble function. At
this point, there are still some non-trivial cosmological scenarios
worth noticing. For example, for constant $G$ the equation above
yields $\dot{\rho}_\Lambda+\dot{\rho}+3H(\rho+p)=0$, which implies
matter non-conservation, whereas if $G$ is variable and matter is
covariantly conserved, i.e. $\dot{\rho}+3H(\rho+p)=0$, then
$(\rho+\rho_\Lambda)\,\dot{G}+G\,\dot{\rho}_\Lambda=0$.  It should
be noticed that, in all these cases, we have the possibility of a
time-evolving cosmological term, $\dot{\rho}_\Lambda\neq 0$.

However, a practical issue is: do we have enough information to
solve the cosmological equations in these variable $\CC$ scenarios?
In other words, can we solve for the fundamental set of variables
\ $H(t)$, $\rho(t)$, $p(t)$, $G(t)$, $\rho_\Lambda(t)$? If we take
any of
these models and combine the corresponding covariant conservation
laws with Friedmann's equation
\begin{equation}
H^{2}=\frac{8\pi\,G }{3}%
\left( \rho +\rho_\Lambda\right) -\frac{k}{a^{2}}\,,  \label{FL1}
\end{equation}
it turns out that we cannot solve the cosmological problem yet. For,
apart from the matter equation of state relating $p$ and $\rho$,
i.e. $p=\omega_m\,\rho$, we still need an additional equation to
solve for the fundamental set of variables. Usually, the additional
equation is provided in the literature on a purely \textit{ad hoc}
basis, see e.g. \cite{Overduin} and the numerous references therein.
In these kind of phenomenological frameworks, one usually starts
with a time evolution equation of the form
\begin{equation}\label{dLdt}
\frac{d\rL}{dt}=F(H,\rho,\rL,...)\,,
\end{equation}
in which $F$ is some prescribed functional of the cosmological
parameters. A prototype approach of this kind was considered e.g. in
\cite{ReutWet87}\,\footnote{For recent developments on general
models with variable cosmological parameters, see
\,\cite{JSHSMPL05}}.

However, it is possible to motivate the needed equation for the CC
evolution in more fundamental terms, namely from the QFT point of
view, and more specifically from the renormalization group (RG)
method \cite{cosm}-\cite{JSHSPL05}. In this case, the aim is to
provide a differential relation (renormalization group equation) of
the form
\begin{equation}\label{RGCC}
\frac{d\rL}{d\ln\mu}=\beta_{\La}(P, \mu)\,,
\end{equation}
where $\beta_{\La}$ is a function of the parameters $P$ of the
general action (\ref{total}) and $\mu$ is a dimensional scale. The
appearance of this arbitrary mass scale is characteristic of the
renormalization procedure in QFT owing to the breaking of scale
invariance by quantum effects. The quantity $\rL$ in (\ref{RGCC}) is
a ($\mu$-dependent) renormalized part of the complete QFT structure
of the CC. Depending on the renormalization scheme, the scale $\mu$
can have a more or less transparent physical meaning, but the
physics should be completely independent of it. Such (overall)
$\mu$-independence of the observable quantities is actually the main
message of the RG; but, remarkably enough, this same message also
tells us that the RG can help to uncover the quantum effects
precisely by focusing on the various $\mu$-dependent pieces
satisfying renormalization group equations like (\ref{RGCC}).
Clearly, equations (\ref{dLdt}) and (\ref{RGCC}) are of very
different nature. We shall further dwell on the (less obvious)
physical interpretation of (\ref{RGCC}) in subsequent sections.

Essential for the RG method in cosmology is to understand that, in
order for the vacuum energy to acquire dynamical properties, we need
an evolving external metric background. For instance, for a
spatially flat FLRW Universe (i.e. $k=0$ in equation (\ref{FL1})) in
the matter dominated epoch, the curvature scalar is given by
\begin{equation}\label{RHHd}
R=-6\left(\,\frac{\ddot{a}}{a}+\frac{\dot{a}^2}{a^2}\,\right)
=-12\,H^2-6\,\dot{H}=-3\,\CC-3\,H^2\,.
\end{equation}
It is, thus, indispensable to have a dynamical background, i.e.
$H=H(t)$, in order to have an evolving $4$-dimensional curvature.
This dynamics (e.g. the observed accelerated expansion) is basically
determined (at low energy) by Einstein's equations (\ref{Einstein}),
but in general we may have the concurrence of the remaining terms in
the extended action (\ref{total}). And, what is more, we also have
the quantum contributions, which can be highly non-trivial and play
a crucial role.

In the absence of a dynamical background, all quantum effects are
static and the RG equation (\ref{RGCC}) does not furnish any
non-trivial information, in the sense that the observed CC value
remains exactly the same at all times. In such situation, there
remains no further $\mu$-dependence apart from the trivial one
associated to the static contributions. Therefore, an equation like
(\ref{RGCC}) is still mathematically possible (cf. section 4), but
we should not be misled by its appearance since it does not reflect
any dynamical effect: it only says that $\rL=\rL(\mu)$ is a static
quantity that can be renormalized at any value of $\mu$ in some
arbitrarily chosen renormalization scheme.

In the general case, the observed CC is obtained from several parts
$\rL^{(i)}(\mu)$ of the renormalized EA, namely from the original
vacuum parameter $\rLV(\mu)$, the possible induced effects
$\rLI(\mu)$ from spontaneous symmetry breaking (cf. section 4) as
well as the contribution from quantum corrections of different
kinds, some of them merely static and some others associated to the
expanding cosmological background. A specific class of these quantum
effects is, ultimately, the potential source of the genuine running
(i.e. of the scaling) of the observed CC with some physical quantity
(cf. section 5). Once the various $\mu$-dependent parts have been
(presumably) been identified with the help of the RG method, the
final value of the CC is generated from the sum of all of them:
\begin{equation}\label{CCi}
\rL^{\rm ph}=\sum_i\,\rL^{(i)}(\mu) \,.
\end{equation}
If this sum could be theoretically computed in QFT in curved
apace-time, it would represent the predicted physical observable
associated to the CC, and could be confronted with the
experimentally determined value $\rL^{\rm exp}\sim 10^{-47}\,GeV^4$.
The number (\ref{CCi}) should be free from all kinds of ambiguities
and, in particular, $\mu$-independent. As for the study of the
scaling properties of this observable, the main point is that not
all of the CC parts of the EA contribute to the physical scaling
with, say, a momentum or the value of an external field (e.g. a
non-trivial metric background). Therefore, the first task to do is
to try to elucidate these especially relevant parts with the help
of the RG. Some clues are given in the next section.

\subsection{What means RG running in Quantum Field Theory}

The properties of the classical theory depend on the dynamical
equations, which can be defined via the principle of minimal action.
So, within the current paradigm, we can state that the definition of
a classical theory starts from establishing its action.

If the asymptotic states can be defined, the quantum theory can be
characterized by the $S$-matrix elements. This is the usual
situation in Particle Physics, for example. However, in gravity,
this is not generally possible and then the definition of the
$S$-matrix is problematic. In this context, the most useful approach
to QFT is accomplished through the notion of effective action (EA),
which is a functional of the mean fields (see, e.g.,
\cite{AbersLee,Coleman85,Brown}). For a QFT with a single scalar
field $\phi$ and Lagrangian density ${\cal L}$, the EA is given by
the functional Legendre transform
\begin{equation}\label{EAdeff}
\Gamma[\phi_c]=W[J]-\int d^4 x \,\sqrt{- g}\ J(x)\,\phi_c (x)\,,
\end{equation}
where $\phi_c$ is the classical or mean field. Here $W$ is a
functional of the source (which in turn is a functional of the mean
field), and is defined through
\beq \,e^{iW[J]} \,=\,\int {\cal D}\phi\, \ e^{iS_{scal}[\phi;
J]}\equiv\int {\cal D}\phi\,\exp\left\{{i}\int d^4x \sqrt{- g}\
\left({\cal L}+J\,\phi\right)\right\}\,, \label{vacEA} \eeq
where $S_{scal}[\phi; J]$ is the classical action of the scalar
field in the presence of the source $J$. The mean field and the
source are Legendre conjugate variables: $\phi_c=\delta W[J]/\delta
J$. Moreover, taking functional differentiations of the EA with
respect to $\phi_c$, one arrives at the one-particle irreducible
Green's functions
\begin{equation}\label{1PI}
\Gamma^{(n)}(x_1,...,x_n)=\frac{\delta^n\Gamma[\phi_c]}{\delta\phi_c(x_1)...\delta\phi_c(x_1)}\,,
\end{equation}
and, finally, to the amplitudes through the reduction
formulae\,\cite{Coleman85}. At the same time, the EA can be seen as
a generalization of the classical action for the quantum domain. It
is important to remember that, in contrast to the classical action,
the EA always has certain ambiguities, which eventually disappear in
the amplitudes. These ambiguities come from the choice of the
parametrization (or gauge fixing, as a particular case) of the
quantum fields and  from the subtraction point associated to the
choice of the renormalization scheme.

The derivation of the EA and working out its ambiguities can be
regarded as the main target of QFT. In the perturbative approach,
one arrives at the corrections (``quantum effects'') to the
classical theory in the form of a power series in the couplings. In
practice, except for some special theories, one can not complete the
task formulated above. The point is that quantum corrections are,
typically, nonlocal expressions which are, at the same time,
non-polynomial in the mean fields. There is, however, a special
sector of the EA where the calculations can be accomplished. In
renormalizable theories, the divergencies of the EA are given by
local expressions which have the same functional dependence on the
mean field as the terms of the classical action have on the
classical fields. These divergencies are eliminated by adding the
local counterterms. After that, one introduces the renormalized
classical action, in which all fields and parameters are functions
of an arbitrary renormalization scale $\mu_R$ (or subtraction
point). Let us clarify that this is a general definition that
applies to any renormalization scheme, so $\mu_R$ here is not meant
necessarily to be the MS mass unit (see below). In more physical
renormalization schemes (for instance, in momentum subtraction
schemes), $\mu_R$ can be a momentum.

The next crucial step is to perform the fundamental identification
between the bare and renormalized theories. To this end, one defines
the renormalized EA to be equal to the bare EA:
 \beq
\,\Ga[g_{\al\be},\Phi(\mu_R),P(\mu_R),\mu_R]=\Ga_0[g_{\al\be},\Phi_0,P_0]\,,
\label{nn82} \eeq
where $\Phi_0$ and $P_0$ stand for the full sets of bare fields and
bare parameters, and $\Phi(\mu_R)$ and $P(\mu_R)$ are the
renormalized ones. As usual, we have the renormalization
transformations between bare and renormalized quantities,
$\Phi_0=Z_{\Phi}^{1/2}\Phi(\mu_R)$ and $P_0=P(\mu_R)+\delta
P(\mu_R)$, where the counterterms are to be fixed by some set of
renormalization conditions (the MS ones being the simplest from the
mathematical point of view). The \textit{identity} (\ref{nn82})
implies that the renormalized action satisfies the renormalization
group equation
\begin{equation}\label{RGEGamm}
\mu_R\,\frac{d}{d\mu_R}\,\Ga[g_{\al\be},\Phi(\mu_R),P(\mu_R),\mu_R]=0\,.
\end{equation}
Although this is a fundamental result, there is no need to prove it,
as some authors curiously attempted to do in a recent
paper\,\cite{AG}, since it is true by definition! The identification
(\ref{nn82}) and corresponding RG equation (\ref{RGEGamm}), holds
for both flat and curved space time, and for all kinds of
renormalizable theories. Writing out the total derivative with the
help of the chain rule one may cast (\ref{RGEGamm}) in the usual
form of a partial differential equation (PDE) satisfied by the
EA\,\footnote{Here we disregard the classical dimension of the
parameters.},
\begin{equation}
\left\{ \mu_R \frac{\partial }{\partial \mu_R }+\beta _{P}
\,\frac{\partial }{\partial P} + \gamma_{\Phi} \,\int
d^{4}x\sqrt{-g}\;\Phi\, \frac{\delta }{\delta \Phi}\right\} \,
\Gamma \left[ g_{\alpha \beta}, P(\mu_R),\Phi(\mu_R),\mu_R \right]
=0\,,\label{RGGamma}
\end{equation}
where an implicit sum over parameters and fields is understood on
the \textit{l.h.s.} of (\ref{RGGamma}). In this expression, we have
the $\beta_P$-functions for all the renormalized parameters of the
theory, and also the $\gamma_{\Phi}$-functions (anomalous
dimensions) for all the matter (mean) fields $\Phi$ (with wave
function renormalization constants $Z_{\Phi}^{1/2}$). They are
defined in the usual way:
\begin{equation}\label{betagamma}
\beta _{P} = \mu_R\frac{\partial P}{\partial\mu_R}\,,\ \ \ \ \ \,
\gamma_{\Phi}=\mu_R\frac{\partial \ln
Z_{\Phi}^{1/2}}{\partial\mu_R}\,.
\end{equation}
The characteristics of the PDE (\ref{RGGamma}) are well-known to be
the running charges $P(\tau)$ for each renormalized parameter, with
$\tau=\ln(\mu_R'/\mu_R)$.

Let us now concentrate for a while on the MS-scheme based
renormalization group with dimensional regularization, as this is
the framework which is more often used in practice, but at the same
time the one whose physical interpretation is less obvious. Here the
renormalization scale is the floating 't Hooft mass unit $\mu$. For
instance, the dimensionally regularized vacuum part of the action
(\ref{total}) in flat space-time boils down to just the bare CC
term,
\begin{equation}\label{dimrLO}
S_{\CC}=-\int d^n x\,\rLO=- \mu^{n-4}\,\int d^n
x\,\left(\rLV+\delta\rLV\right)\,,
\end{equation}
where we have decomposed $\rLO$ (the bare vacuum density in
$n$-dimensions) into the renormalized one $\rLV=\rLV(\mu)$ plus a
counterterm. The MS counterterm  $\delta\rLV$ will be fixed in
section 4. At the moment, this simple example is only to illustrate
that the artificial mass unit $\mu$ is introduced to make the
regularized action dimensionless in such a way that the renormalized
energy density $\rLV(\mu)$ can stay $4$-dimensional\,\footnote{The
scale $\mu$ is the only arbitrary mass unit present in the MS scheme
with dimensional regularization. However, the latter regularization
can also be used in more physical schemes, like in the momentum
subtraction scheme, and then both units $\mu$ and $\mu_R$ are
present, where $|q|=\mu_R$ is the arbitrary momentum subtraction
point.}. In general, the dimensional regularization scale $\mu$
allows that all parameters keep their original dimensions
independent of the regularor $n$.

Similarly, we proceed to make the fundamental identification of the
general form (\ref{nn82}), but in this case for the particular MS
scheme in dimensional regularization:
 \beq
\,\Ga[g_{\al\be},\Phi(\mu),P(\mu),n,\mu] =
\Ga_0[g_{\al\be},\Phi_0,P_0]\,. \label{nn8} \eeq
Again, the overall MS-renormalized effective action $\Ga$ on the
\textit{l.h.s} of this equation does not depend on the floating mass
scale $\mu$. In other words, despite the non-trivial functional
$\mu$-dependence that the various terms of the EA may exhibit (see,
for instance, the QED one-loop example (\ref{QED}) below), the full
renormalized effective action has a value which does not depend on
the particular numerical choice that we make for $\mu$. It means, of
course, that it satisfies a RG equation similar to (\ref{RGEGamm}),
where in this case $\mu_R$ in that equation is replaced by the 't
Hooft mass unit $\mu$.

Let us repeat that the $\mu$-dependence, even though artificial,
manifests itself in the following two important ways:

$\bullet$ \textit{Type-1}\, $\mu$-dependence: All renormalized
fields and parameters depend on $\mu$, i.e. $\Phi=\Phi(\mu)$ and
$P=P(\mu)$. This leads to an implicit $\mu$-dependence of the EA;

$\bullet$ \textit{Type-2}\, $\mu$-dependence: The functional form of
the EA depends explicitly on $\mu$, as also indicated on the
{\textit{l.h.s.}} of (\ref{nn8}).

The combination of these two types of $\mu$-dependencies always
cancels perfectly in the EA, i.e. we have the symbolic annihilation
law\,\footnote{Notice that, in the case of the S-matrix, the
situation is similar, except that there are no $\mu$-dependencies
from the fields $\Phi(\mu)$  because the S-matrix does not depend at
all on them.}
\begin{equation}\label{Type120}
``\,\textit{Type-1}+\textit{Type-2}\,=\,0\,'' \ \
\Longleftrightarrow\ \ \ (\text{``$\mu$-annihilation''}) \ \
\Longleftrightarrow\ \ (\text{RG-invariance})\,.
\end{equation}
This statement is true to any loop order, for otherwise the
renormalized EA would be $\mu$-dependent, violating its own
definition (\ref{nn8}). So, if we take together the two types of
$\mu$-dependencies, we just eliminate entirely the possibility to
use it as a tool to explore the structure of the quantum corrections
-- see the QED case (\ref{QED}) studied below. The use of the RG,
therefore, \textit{requires} that the $\mu$-dependencies of
\textit{Type-1} and \textit{Type-2} are used separately. All these
considerations are of course very well-known by the experts in the
field; we are elaborating on them here because they are at the very
root of some recent misuses of the RG in the literature,
particularly in the cosmological context\,\cite{AG} (see section 4).

The property (\ref{Type120}) is of course valid in any RG framework.
However, in the MS-based scheme, where the renormalized quantities
are simply defined by removing the poles (sometimes along with some
finite terms) of the dimensionally regularized ones, it can be more
difficult to interpret the physical sense of the formers. Moreover,
being the 't Hooft mass unit $\mu$ such an artificial parameter, its
physical interpretation always requires an additional effort. In QFT
in flat space-time, we meet two different standard interpretations
and each of them implies some relation between the MS scheme and
another, more physical, renormalization scheme in the corresponding
limit. Namely, for the $\mu$-dependence of the \textit{Type-1}, we
have to seek correspondence with the momentum subtraction scheme (in
particular, the on-shell scheme whenever possible), which is most
useful one for the analysis of the scattering amplitudes at high
energies\,\cite{Coleman85}. In contrast, the $\mu$-dependence of
\textit{Type-2} is mainly used for the analysis of phase
transitions. Here we associate $\mu$ with the mean value of the
almost static (usually scalar) field and arrive at the
interpretation of the $\mu$-dependent effective potential and its
applications \cite{Sher89}.

Once the divergencies are removed by the renormalization
transformation of fields and parameters, as previously indicated,
the MS-renormalized EA becomes finite. However, the different parts
of the EA became dependent on the arbitrary renormalization
parameter $\mu$. This dependency shows up only in those parts of the
EA which are related to divergences. For example, in massless QED
(which, in practice, means in the limit $|q^2|\gg m^2$), the
electromagnetic part of the one-loop renormalized effective action
has the form
\beq \Ga^{(1)}_{em}\,=\,-\frac{1}{4e^2(\mu)}\, \int
d^4x\,F_{\mu\nu}\,\Big[\,1 -  \frac{e^2(\mu)}{12\pi}\,\ln
\Big(-\frac{\Box}{\mu^2}\Big) \Big]\,F^{\mu\nu}\,, \label{QED} \eeq
where $e(\mu)$ is the renormalized QED charge in the MS scheme. It
is apparent from (\ref{QED}) that the effective (or ``running'') QED
charge in momentum space satisfies
\begin{equation}\label{renormchargeMSQEDHE}
\frac{1}{e^2(q^2)}=\frac{1}{e^2(\mu^2)} - \frac{1}{12\pi^2}\,
\ln\left(\frac{|q^2|}{\mu^2}\right)\,,\ \ \ \ \ \ (|q^2|\gg
m_e^2)\,.
\end{equation}
This relation also follows upon integrating the differential
equation
\begin{equation}\label{RGEQED}
\frac{d\,e(\mu)}{d\ln\mu}=\frac{e^3}{12\,\pi^2}\equiv
\beta_{e}^{(1)}
\end{equation}
from $\mu$ to ${\mu'}$ and then replacing ${\mu'}^2\to |q^2|$.  Here
$\beta_{e}^{(1)}$ is the $\beta$-function of QED at one-loop in the
MS. Equation (\ref{RGEQED}) is correspondingly called the RG
equation of the renormalized charge in the MS scheme. It resembles
equation (\ref{RGCC}) for the $\CC$ term (on which we will return
later). Notice from (\ref{renormchargeMSQEDHE}) that $e(q^2)$
increases with $|q^2|$, as expected from the non-asymptotically free
character of QED -- a well tested feature of this theory.

The remarkable property of tracking explicitly the $\mu$-dependence
of the various parts of the EA is that the high-energy limit of the
theory can be correctly reproduced, in the leading approximation, by
taking the limit of large $\mu\gg m$. It is understood, when
performing this limit, that all of the degrees of freedom, say
$N_f$, having e.m. charge $e$ and whose masses satisfy $m_f<\mu$,
are included in the calculation of the $\be_e$-function. In
practice, this means that we can set the correspondence $\mu\to
|{q'}|$, and in this way the QED running charge at the energy scale
$|q|$ is related to the corresponding value at the other high energy
scale $|{q'}|$ through
\begin{equation}\label{renormchargeMSQEDHE2}
e^2({q'}^2)=\frac{e^2(q^2)}{1 -N_f\,
\displaystyle\frac{e^2(q^2)}{12\,\pi^2}\,
\ln\frac{|{{q'}}^2|}{|q^2|}}\,,
\end{equation}
a relation which, by the way, can be tested experimentally since it
depends on physical quantities only\,\cite{PDG06}.

Due to the simplicity of the form factor for the massless case (the
above expressions are a nice illustration) one can always restore
such form factor using the $\mu$-dependence, i.e. the solution of
the RG equation (\ref{RGEQED}) in the MS scheme and the
correspondence of $\mu$ with a high energy momentum. Let us clarify,
however, that in QED there is no need to use the MS, if one does not
wish to, because we may naturally proceed in the on-shell scheme,
where the renormalization can be performed by subtracting the
Green's functions on the mass-shells of the physical particles.
Nevertheless, the lessons that we can learn from this simple QED
example in the MS scheme are very important and should be stressed,
to wit:

 \textit{Lesson} \textbf{i)}. The presence of the
dimensional parameter $\mu$ is the characteristic sign of the
dynamical breaking of scale invariance in QFT. As a result, the
$\mu$-dependence of the EA (or of the scattering matrix, if it
exists) can be used in practice as a parametrization of the
different quantum effects, despite that the EA and the scattering
matrix are, overall, $\mu$-independent. Thanks to this
$\mu$-parametrization provided by the MS-scheme based RG, we can tag
the different types of quantum effects that the EA contains and this
provides a clue for their identification, classification and
evaluation.

\textit{Lesson} \textbf{ii)}. If the scattering matrix
can be defined, then at high energy the $\mu$-dependence of the
MS-renormalized amplitudes ($S$-matrix elements) can be translated
into momentum dependence, and this leads to the notion of running
charges with momentum, e.g. $e=e(q^2)$ in QED.

Keeping in mind these lessons is important, because it is often not
possible to use a physical renormalization scheme. In Particle
Physics, we happen to meet the two situations, namely we have QED
and the electroweak standard model ($SU(2)_L\times U_Y(1)$) as
excellent prototypes of theories where the on-shell scheme is
ideally applicable, but we also have the archetypal example of
quantum field theory where an off-shell renormalization scheme is
mandatory: QCD, the $SU(3)_c$ gauge theory of strong interactions.
In practice, this means that in QCD we must use some off-shell
momentum subtraction scheme, or most often, simply the MS
scheme\,\cite{Coleman85,Brown}. It turns out that, in the gravity
framework, we have a kind of similar (actually more complicated)
situation when dealing with the CC parameter, in the sense that
it is technically impossible at the moment to examine
the renormalization of $\CC$ in a physical subtraction scheme, even
though it would be most desirable. As a result, we are temporally
forced to use the MS scheme and examine the $\mu$-dependence of the
various terms that contribute to the vacuum energy density, as a
modest tool to dig out the possible quantum effects of the theory,
in particular the scaling properties of the CC (see sections 4 and 5
for more details).

In the massive case, the situation is more complicated. In the high
energy limit there is an effective coincidence with the MS-based
result, so the $\mu$-dependence provides reliable information. At
the same time, the low-energy limit has, typically,  no direct
relation to the $\mu$-dependence. This is well-known in QED, for
example, where the low energy limit of the vacuum polarization
tensor is characterized by a behavior of the form $\Pi(q^2)\sim
q^2/m^2$\,\cite{Coleman85}. Indeed, in this regime, the masses of
the quantum fields dominate in the internal lines of the loops and
in many cases we observe the quadratic decoupling, according to the
standard Appelquist and Carazzone theorem \cite{AC}. In such
situation, one has to be very careful in applying the MS-based
results. In particular, the $\mu$-dependence is no longer a good
parametrization of these effects. The simplest option is to assume a
``sharp cut-off'', that means to disregard completely the
contributions of massive fields at the energy scale below their mass
($|q^2|<m^2$) and, at the same time, treat their contributions above
the proper mass scale ($|q^2|>m^2$) as high-energy ones, without
taking decoupling effects into account. This procedure has been
reflected in our QED example above. Let us notice that this was also
the precise recipe adopted in our first articles on the CC
running\,\cite{cosm,nova}, while in subsequent papers on the subject
\cite{babic,CCfit,Gruni} the starting hypothesis was a quadratic
decoupling. We shall discuss both options in more detail in the next
sections.

At this point, it is appropriate to remark that
\textit{Lesson}\textbf{ ii)} above, although it is of invaluable
help in Particle Physics, it has a limited scope for the
gravitational applications. In practice, due to the simplicity of
the MS scheme, one often tries to use it as a heuristic approach to
extract information on the radiative corrections. However, the basic
problem that we have here is that, in the presence of a
non-vanishing $\CC$, the space-time can not be flat -- as it is
transparent from Einstein's equations (\ref{Einstein}). Notice that
even if the $\rL^0$ term in (\ref{total}) would be set to zero, it
is impossible to have a QFT in strict flat space-time since
Einstein's equations could not be satisfied. In fact, take the
simplest case of a free scalar field. The closed one-loop diagrams
would generate the following (bare) vacuum energy density
contribution in dimensional regularization:
\begin{equation}\label{rhovac}
{\bar V}_{vac}^{(1)}=\sum_k \frac12 \hbar\, \omega_k \equiv
\frac12\,\mu^{4-n}\, \int\frac{d^{n-1}
k}{(2\pi)^{n-1}}\,\hbar\,\sqrt{\vec{k}^2+m^2}\,.
\end{equation}
Here we have included $\hbar$ for a while, just to {emphasize} that
(\ref{rhovac}) acts as an effective CC induced by the quantum
theory. In order to satisfy Einstein's equations, the space-time
geometry on the \textit{l.h.s.} of Eq.\,(\ref{Einstein}) is bound to
become curved appropriately as a backreaction to the energy input on
its \textit{r.h.s.}. Furthermore, the subsequent renormalization of
this QFT contribution requires to have the bare term $\rL^0$ back in
the original action (see section 4). So, within the context of QFT,
the geometry of the space-time can never be flat, strictly speaking.

The previous remark is important, for if one uses the MS scheme to
renormalize the action (\ref{total}) the resulting $\mu$-dependence
of the various terms can be hardly associated to any $q$-dependence
of a scattering amplitude (in contrast to QED or QCD). The reason is
that the $S$-matrix cannot be properly defined in a curved
background. Furthermore, except for the simplest static spaces, the
definition of the effective potential problem is not easy in curved
spaces \cite{sponta}. And, finally, in curved space the existence of
an almost static scalar field does not imply, in general, a static
metric or static curvature. This is typical of a theory where two
fields are present, in which the heavy one can be a scalar and the
massless one represents the metric excitations.

Notwithstanding the limited applicability of \textit{Lesson}
\textbf{ii)} in the gravitational domain, let us emphasize that
\textit{Lesson} \textbf{i)} is, in contrast, still perfectly valid.
As a matter of fact, it will be our main guiding paradigm in our
search for the ultimate physical sources of the CC running within
QFT in curved space-time. It actually gives the clue for the
interpretation of the RG equation (\ref{RGCC}) for $\CC$. Namely, by
integrating this equation with respect to $\mu$, one obtains the
functional $\mu$-dependence $\rL=\rL(\mu)$ associated to that
particular part of the EA  contributing to the observed CC, and by
an appropriate choice of $\mu$ in the cosmological context one may
estimate the numerical contribution from this part.

Let us insist that there is no such thing as an RG equation
(\ref{RGCC}) for the full physical CC, since the latter -- being an
observable quantity -- does not depend on $\mu$, cf.
Eq.\,(\ref{CCi})! This trivial, but relevant, observation shows the
radical distinction between the kind of equations (\ref{dLdt}) and
(\ref{RGCC}) introduced in section 2. In short, the renormalization
group Eq.\,(\ref{RGCC}) can only be applied to some CC parts of the
EA which are parameterized by $\mu$. The real problem here, of
course, is that we don't know all the parts of the EA that feed the
entire physical CC (most conspicuously the dynamical ones associated
to the expanding background). Therefore, the particular choice of
$\mu$ (for example, $\mu=H, \,1/a,$ etc., see the next sections) can
be relevant for the numerical estimate of the contribution from a
concrete $\mu$-dependent part that is known.

\section{Renormalization group for $\CC$:  brief history}

Let us present a brief - most likely incomplete - list of the
publications devoted to the renormalization group for the CC. A
similar classification has been considered in a recent general
review on the CC problem \cite{nob}, but our comments may be
somewhat different from the ones given in that review and are
concentrated almost exclusively on the RG approach.

We start our list from the papers that treated the renormalization
group for the CC in a mathematically consistent way, but did not
consider cosmological applications. For example, in the pioneering
paper by Nelson and Panangaden \cite{NelPan}, the renormalization
group for the CC and other parameters of the vacuum action was first
linked to the global scaling of the metric. The same ideas were
developed in more formal works by Buchbinder \cite{buch84} and Toms
and Parker \cite{Toms83}. It is also worth mentioning the first
practical calculations \cite{buchod-82}. A consistent pedagogical
account of the renormalization group in curved space, together with
much more references on the subject, can be found in the book
\cite{book} -- the review of some recent achievements has been
presented in \cite{PoImpo}. In the above mentioned papers, gravity
was treated as a classical background and only matter fields were
quantized.

Other important works where the renormalization group for the CC has
been considered are the papers by Salam and Strathdee \cite{Salam}
and by Fradkin and Tseytlin \cite{frts82} (see also the subsequent
calculations of \cite{avbar,Gauss}). The object of quantization in
these papers was higher derivative quantum gravity, and its
renormalizability enables one to use the standard MS-scheme based
renormalization group for $\CC$, $G$ and other parameters. Let us
note, for the sake of completeness, that \cite{frts82} includes also
an alternative (on shell) approach to the renormalization group
equation for $\CC$, which can be applied even to the
non-renormalizable quantum General Relativity.


\subsection{AI-type approaches}

In the 1982 review paper \cite{Pol81}, Polyakov made the intriguing
observation that the IR effects of the unknown quantum gravity
theory might lead to the effective screening of the CC at the cosmic
scale. We can consider this remark as a starting point of the {\it
AI} approaches (following the notation of section 1). In 1990 Taylor
and Veneziano discussed the renormalization group as an instrument
for solving the CC problem \cite{TV}. The main difficulty with these
approaches is that no working
model leading to the CC suppression in the infrared (IR) has been
found.

The first such model traces back to 1992, which is when Antoniadis
and Mottola introduced the idea that the ``IR quantum gravity''
could solve the CC problem through the mechanism of RG screening
\cite{antmot} (see also \cite{AMM} for further developments on this
proposal, and \cite{Prokopec07} for more recent discussions). This
model of quantum gravity has been suggested earlier in
\cite{odsh91}, it implies that only the conformal factor of the
metric should be quantized and that the starting action must be the
one induced by the conformal anomaly \cite{rie}. The effective IR
screening really takes place in this model, so in this sense it is
very successful. However, it is by far not complete, because there
are three unclear aspects, namely: \ {\it (i)} It is implicitly
assumed that the IR decoupling of the massive mode of the conformal
factor does not occur at low energies;  {\it (ii)} Why we should
quantize only the conformal factor of the metric? In particular, the
problem of higher derivative ghosts is present in the ``IR quantum
gravity'', exactly as in the usual higher derivative quantum gravity
\cite{stelle}; {\it (iii)} The ambiguities which are typical for the
RG for CC in the higher derivative quantum gravity \cite{frts82},
should also be present in the theory of \cite{antmot}. All three
points are difficult to address and this has been never done in a
comprehensive way, despite that an attempt to deal with the item
{\it (i)} has been done in \cite{Cognola}.

Another two realizations of the ``IR quantum gravity'' and the
screening of CC can be attributed to the works \cite{reuter} and
\cite{woodard}. In the first case, the central idea is to assume the
existence of the non-Gaussian fixed point in quantum gravity. After
that, the existing ambiguity in the non-perturbative RG equations
for the CC and the Newton constant $G$ is fixed by requiring that
$G$ tends to the constant value in the IR limit. As a result, we
observe a screening of the CC at low energies. Let us remark that
exactly the same idea has been published earlier in the paper
\cite{lam}, which belongs to the {\it AII} class of approaches and
will be discussed in the next subsection. In an alternative approach
which was used in \cite{woodard}, the IR screening of the CC is
derived without the explicit use of RG. The calculations are based
on the original coordinate-representation Feynman diagrams in the
deSitter space (dS). Unfortunately the calculations on dS do not
distinguish the CC, Einstein-Hilbert term and the higher derivative
terms in the vacuum action, while in reality all of them have their
own RG \cite{PoImpo} and hence their own leading-log corrections.
Therefore it is unclear what is actually screened in this model.
This shortcoming is indeed still present in the last version on the
model \cite{woodard-RG}, despite it is stated that this approach is
more correct than the RG one.

\subsection{AII and BII-type approaches}

The first model with the IR screening for the CC due to the quantum
effects of matter fields was {constructed by one of the present
authors in \cite{lam} (see also further discussion in \cite{Irgac}
and similar model suggested by Jackiw et al} \cite{jackiw}). Let us
remind that the observable CC consists of the vacuum and induced
parts. The vacuum CC is necessary in the quantum field theory, since
it provides renormalizability in the vacuum sector (see \cite{nova}
for a detailed discussion). However, if the theory is completely
massless, one can avoid introducing the vacuum CC and then one needs
to have the screening for the induced component only. The main
assumption in \cite{lam} was that the origin of all masses is the
Coleman-Weinberg mechanism (dimensional transmutation) within some
GUT model, where the scalar field $\phi$ is non-minimally coupled to
curvature (i.e. $\sim\xi\,\phi^2\,R$) and has self-interaction
($\sim f\,\phi^4$). The induced quantities are, in this case,
\beq (16\pi G)^{-1} \,\,\sim\,\, <\xi \phi^2> \qquad \mbox{and}
\qquad \rho_\La\,\, \sim\,\, \,<f \phi^4>\,, \label{induce} \eeq
The RG behavior of these induced quantities depend on the RG
equations for $\,\phi,\,\,f,\,\,\xi$. Fixing the usual ambiguity in
the $\ga$-function for $\phi$ such that $G \to const$ in the IR, for
a wide class of GUT models we arrive at the effective screening of
the CC,
\beq \rho_\Lambda \,\,\sim \,\,
\left( \mu /\mu_0  
\right)^{2\left|A\right|g^2} \longrightarrow 0\,, \eeq
where $\mu$ and $\mu_0$ are the current cosmic energy scale and
initial high energy scale, respectively.

Despite this model does not look phenomenologically realistic, it
shows that the quantum effects may be relevant for solving the CC
problem, so that it is worthwhile to work further in this direction.

The first {\it BII}-type model for the study of the CC running has
been suggested by the present authors in \cite{cosm,nova}, where we
used the standard MS-based RG for the CC
\cite{NelPan,buch84,Toms83}\,\footnote{See e.g. \cite{Brown} for a
pedagogical introduction to the RG for the CC in the MS scheme.}.
The scale parameter $\mu$ has been associated to the energy scale
defined by the critical density, i.e. $\mu\sim\rho_c^{1/4}$. During
the hot stages of the evolution, say at temperature $T$, such scale
obviously behaves as $\mu\sim T\sim a^{-1}$, where $a=a(t)$ is the
scale factor, whereas in the current Universe it behaves as $\mu\sim
\sqrt{H_0\,M_P}$, i.e. as the geometric mean of the two extreme
physical energy scales in the IR and the UV in our Universe. The use
of the MS-scheme based RG was implemented through the ``sharp
cut-off'' approximation (see section 2.3). In this case it means
that, in the current Universe, we assumed that there is a sharp
decoupling of all particles having masses $m$ above the
$\mu\sim\rho_c^{1/4}\sim\sqrt{H_0\,M_P}$ energy scale. At the same
time, following the MS-based RG prescription, we took into account
the contribution from all the light degrees of freedom satisfying
$m<\mu$ and considered their effect for the running of all the
parameters, without bothering about their mass effects (as this is
part of the MS prescription). Since the lightest neutrinos could
have masses below the current value of the cosmic scale in this
framework, i.e. $m_{\nu}\lesssim\sim\sqrt{H_0\,M_P}\sim
10^{-3}\,eV$, our procedure took into account the quantum effects on
the CC running from the lightest neutrinos, which are the natural
candidates here from the SM, including some hypothetical light
scalar fields which may emerge in the extensions of the SM (like
axions and cosmons\,\cite{PSW}). As a result, the naturally
predicted value of the CC evolution in this range is
$\Delta\rL\sim\mu^4\sim\,H_0^2\,M_P^2$, which is of the order of the
CC value itself at the present time. The cosmological implications
based on this RG scenario have been developed in
\cite{GHScosm,Bauer1}, and also recently in an analysis of the
cosmological neutrino mass bounds in the presence of a running CC,
see \cite{Bauer2}\,\footnote{For other interesting implications of
the neutrino physics on the running of the CC, see
\cite{Peccei04}.}.

The subsequent works on the {\it BII} approach have explored the
effects of a smooth Appelquist and Carazzone - like decoupling and
greatly benefited from the contributions of Babic, Guberina, Horvat
and \v{S}tefan\v{c}i\'{c}\,\cite{babic,Babic2}. An interesting
proposal was the possibility to use the Hubble scale ``$\mu=H$''
(cf. \,\cite{nova}) along the idea of ``soft decoupling'' of the GUT
fields. This scenario provided a plausible phenomenological model
for the physical running of the CC in a mass-dependent framework,
and was amply explored in subsequent
publications\,\cite{CCfit,IRGA03,Gruni,JSHSPL05,FSS1}\,\footnote{Let
us note that the two scale associations $\mu \sim 1/a$ and $\mu \sim
H$ could be related \cite{fossil}.}. It is based on the IR ansatz:
\begin{eqnarray}
\label{expecVERen}
\rL(H)=\rL^0+\frac{3\nu}{8\pi}\,\left(H^2-H_0^2\right)\,M_P^2 +{\cal
O}\Big(H^4\Big)+{\cal O}\Big(\frac{H^6}{m^2}\Big)+...\,,
\end{eqnarray}
in which the leading term (the so-called ``soft-decoupling'' term)
is quadratic in the expansion rate $H$, and the remaining powers of
$H$ are the ordinary decoupling contributions of the
Appelquist-Carazzone type\,\cite{AC}. Here $\nu$ is a small
parameter, essentially given by the ratio squared of the average
mass $M$ of the GUT particles to the Planck mass, specifically
$\nu=({1}/{12\pi}){M^2}/{M_P^2}$. In Ref.\,\cite{FSS1}, this
parameter was stringently bounded ($\nu < 10^{-4}$) from the
analysis of cosmological perturbations. Notice that, in this model,
one also obtains that the typical amount of running is
$\Delta\rL\sim\,H^2\,M_P^2$, as in the previous one. However, in the
present case, the concept is different, namely, the evolution of the
CC is not  $\sim m^4$ (with $m\lesssim\mu\sim \sqrt{H\,M_P}$), but
of the soft decoupling type $q^2\,M^2$, where $|q|^2\sim |R|\sim
H^2$ is the typical momentum squared of the cosmological gravitons
-- see (\ref{RHHd}) -- and involves, in contrast, the heaviest
degrees of freedom. The explicit appearance of masses and momenta
makes this formulation more physical and the association ``$\mu=H$''
can be interpreted as a momentum subtraction scale, i.e.
$\mu_R=|q|\simeq H$ (quite different from the MS case) along the
lines of (\ref{nn82}).

Let us clarify that the ansatz (\ref{expecVERen}) refers to the
effective IR behavior, assumed in these references, for the vacuum
part of the CC only. In order to really build up constructively this
expression using the RG method, we should go through the computation
of the various quantum effects. Let us notice that this approach
could emerge as a most natural description of the vacuum effective
action at low energies \cite{nova,fossil,PoImpo}, as we shall also
discuss in section 5.

Finally, we have to mention the explicit derivation of the
Appelquist and Carazzone theorem for gravity \cite{apco}. The net
result of these calculations is that we can establish and prove this
theorem for the higher derivative sector of the vacuum action, but
not for the CC and Einstein-Hilbert terms. It is important to
emphasize that this does not mean that there is no running in these
sectors. The output of the calculations of \cite{apco} only shows
that the currently available methods are not appropriate for the
mentioned purpose. These methods are based on the flat space
expansion and can not lead to the desired running of $\rho_\La$ and
$G$ independently of whether such running takes place or not.

\section{Effective potential and cosmological constant}

The idea that the cosmological term could be adjusted dynamically
first appeared, historically, within the context of the effective
potential models of scalar field theories modeling the vacuum
state\,\cite{Dolgov82,PSW,Adjusting}. More recently, the scalar
field models have overflowed the ``cosmological marked'' in the form
of quintessence, k-essence and the like\,\cite{DEquint} and have
strengthened the perspective for a possible dynamical nature of the
dark energy. Unfortunately, at the moment, all the proposed scalar
fields that hypothetically could solve the CC problem are completely
\textit{ad hoc} and, in addition, none of them has any obvious
relationship with the Particle Physics world, e.g. with the SM of
the strong and electroweak interactions or with some promising
extension of it (like the MSSM, or some favorite supersymmetric
GUT\,\cite{MSSM}).

On the other hand, in the renormalization group context, a first
useful proposal to describe the dynamical dark energy as a running
CC was put forward in \cite{cosm,nova} within the context of the SM
of Particle Physics, and it made use of the properties of the
effective potential of the SM (i.e. the Higgs potential). Later on,
in \cite{CCfit,IRGA03,Gruni,JSHSPL05} various cosmological models
have been developed on the basis of the possible running of the CC
along the lines sketched in section 2.2 and, in particular, using
the phenomenological ansatz (\ref{expecVERen}).

{Instead of thinking about the cosmological models with running
$\CC$ one may adopt, instead, an opposite attitude and try to prove
that such running is mathematically impossible.} An attempt in this
direction has been undertaken in a recent {article} \cite{AG}.
Unfortunately this short, simple and apparently clear work is
plagued by a number of conceptual mistakes. After this preprint was
submitted to the arXiv, several colleagues asked us to explain the
situation\,\footnote{Let us remark, in passing, that the version of
this paper which has been made public had no essential difference
with the one which these authors have sent to us prior submission.
We have explained them what is the correct point of view, but the
only effect of these explanations on them was to thank us for the
correspondence in the Acknowledgments of \cite{AG}.}. So, correcting
the misconceptions of \cite{AG} seems a necessary step to restore
the truth.

The consideration of \cite{AG} is based on the renormalization group
equation for the effective potential of the real massive scalar
field with the $\la\phi^4$ interaction. The classical part of the
potential is
\beq U(\phi) = \frac{m^2 \phi^2}{2}+\frac{\la \phi^4}{4!}\,.
\label{clpo} \eeq
The quantum corrections are then computed in the MS scheme with
dimensional regularization. One obtains an expression of the form
\begin{equation}\label{Vloops}
\EPR(\phi)=U(\phi)+\hbar\,{V}^{(1)}(\phi)+\hbar^2\,{V}^{(2)}(\phi)+...,,
\end{equation}
where the powers of $\hbar$ count the number of loops. Later on in
this section we shall consider the specific case of the one-loop
correction.

This situation applies also to the Higgs potential of the SM, which
was addressed in the early
papers\,\cite{cosm,nova}. In these papers, we defined (in some
equivalent notation) the expression
\begin{equation}\label{Lambdamu}
\rL(\mu)=\rLV(\mu)+\rLI(\mu)\,,
\end{equation}
which is the sum of the vacuum and induced parts of the CC. Of
course, the quantity (\ref{Lambdamu}) is not intended to be the
entire observable CC, i.e. Eq.\,(\ref{CCi}), because the above sum
is $\mu$-dependent whereas the observable CC is not! The two terms
on the \text{r.h.s.} of this expression constitute only the,
so-called, \textit{Type-1} contributions (see section 2.3) to the
vacuum and induced parts of the CC. In other words, they can be
regarded as two independent contributions to the overall sum on the
\textit{r.h.s.} of Eq.\,(\ref{CCi}). As we remarked in section 2.3,
the RG method is based on tagging and classifying the two types of
contributions (\textit{Type-1} and \textit{Type-2}) and it is
essential not to mix them, otherwise the $\mu$-dependence disappears
altogether, as indicated in (\ref{Type120}); in this case, we would
lose the chance to pinpoint where are the possible parts of the EA
connected with the physical running. Below we explain the meaning of
(\ref{Lambdamu}) in more detail and we identify explicitly  the two
terms $\rLV(\mu)$ and $\rLI(\mu)$.

The vacuum CC for the above model can be included into the scalar
potential in the form $hm^4$, where $m$ is the scalar mass and $h$
is an independent dimensionless parameter, similar to the parameter
$f$ in (\ref{CCd1}). Let us recall that the effective potential for
local field theories is related to the effective action in the limit
of constant mean field ($\phi=const.$) through
\begin{equation}\label{VeffEA1}
\Gamma[\phi]=-\int d^4 x\,\sqrt{-g}\ \EPR
(\phi,m^2,\la,h,\mu)=-\Omega\,\,\EPR (\phi,m^2,\la,h,\mu)\,,
\end{equation}
where $\Omega$ is the space-time volume. Therefore, from the
fundamental RG equation (\ref{RGGamma}) satisfied by the
renormalized effective action in the MS scheme with dimensional
regularization ($\mu_R\to\mu$), we immediately find the
corresponding RG equation satisfied by the renormalized effective
potential of such theories:
\beq \Big(\mu\frac{\pa}{\pa \mu}+\be_{\la}\frac{\pa}{\pa \la} +
\ga_m m^2 \frac{\pa}{\pa m^2} +  \be_h \frac{\pa}{\pa h}+\ga_\phi
\phi \frac{\pa}{\pa \phi} \Big)\,\EPR(\phi,m^2,\la,h,\mu)\,=\,0\,,
\label{RG} \eeq
where we have specified the contributions from the parameters
$P=\lambda,m,h$, with corresponding $\beta_P$-functions
(\ref{betagamma}) (the one for the mass squared is usually relabeled
as\, $\beta_{m^2}\equiv\gamma_m\,m^2$). As remarked in section 2.3,
there is nothing to be proven here; equation (\ref{RG}) is true just
by definition of effective action and effective potential. As a next
step, the authors of \cite{AG} notice that, in the minimum of the
potential, $\pa V/\pa \phi=0$, and therefore the value of the
potential in the minimum is $\mu$-independent\footnote{One can
immediately notice that in the $\la\phi^4$-theory the use of the
condition  $\pa V/\pa \phi=0$ is not very relevant, because
$\ga_\phi=0$ at one-loop.}. Their conclusion (although obvious
\textit{ab initio!}) is that the minimal value of the overall
effective potential is $\mu$-independent. The aforementioned authors
interpret this result as a kind of ``non-running theorem'', i.e. as
a formal demonstration that the observed value of the vacuum energy
does not run, in contrast to the conclusion of our original papers
\cite{cosm,nova}. The effect of other fields (e.g. the SM
constituents),  or the presence of an external gravitational
background, does not change -- according to the authors of \cite{AG}
--  this conclusion at all.

In the light of the discussion presented in the previous sections,
specifically in  2.2 and  2.3, it should be vastly apparent by now
that the main theses of Ref.\,\cite{AG} are completely unjustified
(in fact, entirely wrong) as they are based on a (severe) conceptual
mistake about the significance of the RG as a tool to explore the
quantum effects in QFT. Let us also note that while these authors
cite only our papers on the CC running, their ``criticisms'' go
simultaneously at the heart of the sizeable body of respectable
works on the CC running in the literature, many of which we have
cited in section 3. In fact, most of them are also based on the
MS-scheme of renormalization\,\footnote{For example, the RG approach
of our first paper \cite{cosm} was exactly a standard MS-based one
as in the works \cite{NelPan,buch84,Toms83}. The important new
element was the physical interpretation of it within the Particle
Physics phenomenological framework, especially within the SM of
strong and electroweak interactions.}.
For better clarity,  let us try to classify in some more detail the
incorrect misconceptions contained in Ref. \cite{AG} (which they
used against our original works\,\cite{cosm,nova}). The main reason
for doing this is to try to ``undo'' as much as possible the chain
of possible confusions that this work might have spread over some
readers. It will also serve as a basis for a pedagogical discussion
of the subject.

\vspace{0.2cm}

\noindent {\Large $\bullet$} First. Physical observables (in
particular the observed CC obtained from supernovae, CMB and LSS
data\,\cite{CCdata}) are RG-invariant (i.e. $\mu$-independent)
quantities. Therefore, trying to prove that a physical observable is
RG-invariant is a (useless) tautology.

\vspace{0.2cm}

\noindent {\Large $\bullet$} Second. The essential confusion of
these authors is to incorrectly identify the physical running of the
cosmological term with the $\mu$-dependence. This is obviously wrong
and is connected with the previous point. One thing is the
mathematical ``running'' of some parameters of the EA with the
floating mass scale $\mu$ (i.e. what we called \textit{Type-1}
$\mu$-dependence in section 2.3), together with the explicit
$\mu$-dependence of some parts of the EA (\textit{Type-2}
$\mu$-dependence), and another (quite different) thing is the
measurable scaling of a physical observable (in this case the
observed CC) with a physical quantity, say a momentum $q$ or an
external field (such as e.g. the metric and its derivatives in the
presence of a non-trivial background). While the $\mu$-dependence
can be used in some cases to trace the $q$-dependence (or the
external field dependence) of the quantum corrections to the
S-matrix (or to the EA), the $\mu$-dependence in itself has no
intrinsic physical meaning, because $\mu$ cancels in the overall
result. The physical running (if it is there at all) is not in $\mu$
but in $q$ (or in the dynamical properties of the external
background metric). For example, in the model of
Eq.\,(\ref{expecVERen}), the physical running lies in the
$H$-dependence of the result. This is of course true irrespective of
whether the floating mass scale $\mu$ of the various pieces is
explicitly kept or not, even though we know it must cancel in the
complete CC expression. Still, it is customary (by an abuse of
language) to refer to this $\mu$-dependence of the renormalized
quantities as a kind of ``running'' with $\mu$. We have no problem
in keeping this usage of words, but the reader should be careful in
distinguishing it from the physical running or scaling of the
parameters with momenta or field strengths.

\vspace{0.2cm}

\noindent {\Large $\bullet$} \ Third. Equation (\ref{RG}) is, in
fact, a sum of two independent equations. Indeed, one can split the
overall effective potential as a sum of two pieces, the
$\phi$-independent (vacuum) term and the $\phi$-dependent (scalar)
term, as follows:
\beq \EPR(\phi,m^2,\la,h,\mu)=V_{scal}(\phi,m^2,\la,\mu) +
V_{vac}(m^2,\la,h,\mu) \label{VeffT}\,. \eeq
In this way we can also split the RG equation (\ref{RG}) into two
independent RG identities:
\beq \Big(\mu\frac{\pa}{\pa \mu} + \be_{\la}\frac{\pa}{\pa \la}
+\ga_m m^2 \frac{\pa}{\pa m^2} +\ga_\phi \phi \frac{\pa}{\pa \phi}
\Big)V_{scal}(\phi,m^2,\la,\mu)\,=\,0\,, \eeq\label{RG-sep-EP}
\beq \hspace{2.5cm}\Big(\mu\frac{\pa}{\pa \mu} + \be_{\la}
\frac{\pa}{\pa \la}+\ga_m m^2 \frac{\pa}{\pa m^2}+\be_h
\frac{\pa}{\pa h}\Big)V_{vac}(m^2,\la,h,\mu)
\,=\,0\,.\qquad\qquad\qquad\qquad\qquad \label{RG-sep} \eeq
In order to understand the origin of this splitting, one has to
introduce the functional called effective action of vacuum. This
object is that part of the overall EA which remains nonzero when the
mean scalar field is set to zero. The vacuum EA comes from the
Legendre transform of the generating functional of the one-particle
irreducible Green's functions without the source term for the scalar
field, i.e. $J=0$ in equations\,(\ref{EAdeff}),(\ref{vacEA}):
\beq e^{i\Ga_{vac}}\,=\,e^{iW_{vac}} \,=\,\int {\cal D}\phi \
e^{iS_{scal}[\phi; J=0]}\,. \label{vacEA2} \eeq
In flat space-time, the functionals $\Ga_{vac}$ and $W_{vac}$ are
equal numbers which do not depend on the field variables; they are
the generators of the proper vacuum-to-vacuum diagrams. In flat
space-time, these functionals are usually disregarded as irrelevant
constants, and moreover their divergences can be eliminated by
changing the operator ordering. At the functional level, this is
equivalent to normalize the functional (\ref{vacEA2}) to one (using
the arbitrary normalization prefactor that carries with it).

However, in the presence of a gravitational background, the
situation can be non-trivial, in contrast to the claims of
\cite{AG}. Due to the presence of an external metric, it is more
advisable to take $\Ga_{vac}$ to be a subject of the renormalization
procedure and remove its divergence by renormalizing the parameter
$h$ (for the sake of convenience of the reader, we keep some of the
notations of \cite{AG} in this section). From the RG-invariance of
the renormalized EA -- see Eq.\,(\ref{RGGamma}) -- it follows
immediately the $\mu$-independence of the renormalized functionals
$W_{vac}$ and $\Ga_{vac}$ and, therefore, we arrive at the second
identity (\ref{RG-sep}) for the vacuum part of the effective
potential, while the first identity is the result of the subtraction
of (\ref{RG-sep}) from (\ref{RG}).

The net result of these considerations is that the vacuum and matter
parts of the EA and, consequently, of the effective potential, are
$\mu$-independent separately and no cancelation between them can be
expected.

\vspace{0.2cm}

\noindent {\Large $\bullet$} Fourth. The quantity (\ref{Lambdamu}),
which we used in \cite{cosm,nova}, may now be unambiguously defined;
namely, it is just the sum of the respective \textit{Type-1}\,
$\mu$-dependent parts of the two RG-invariant pieces, $V_{scal}$ and
$V_{vac}$, of the effective potential. The \textit{Type-2}
dependences from these parts of the effective potential are
\textit{not} included, otherwise the $\mu$-annihilation relation
(\ref{Type120}) would hold.  As a result, $\rL(\mu)$ in
(\ref{Lambdamu}) is a neatly $\mu$-dependent expression;
specifically, it consists of the parts that we would usually put in
correspondence with the momentum subtraction scheme (or even with
the on-shell scheme), if that would be possible. But in those cases
where the correspondence with a physical renormalization scheme is
especially difficult and unclear, one may try the sharp cut-off
approximation in the MS scheme; and this is precisely what we did in
\cite{cosm,nova}.

\vspace{0.2cm}

\noindent If the ``proof'' of the no-running CC proposed in
\cite{AG} would make sense at all, one could immediately extend its
scope with major implications. Thus, one could easily
``demonstrate'' such no-running feature in all of the theories where
the overall $\mu$-dependence is absent, some of them being quite
conspicuous by the way, such as the NJL model, QED and QCD. It is
especially easy to do for the NJL model, because the difference with
the CC case can be disposed of by just changing the parametrization
of the external metric, see equation (\ref{CCd1}) and the
corresponding discussion. Of course, proceeding in this manner would
be a rather unproductive, in fact absurd, way of using the MS-scheme
based RG. Nevertheless, it seems that it is precisely what these
authors have done in their completely unsuccessful analysis of the
CC problem.

\vspace{0.2cm}

\noindent {\Large $\bullet$} \ For a correct analysis, the two RG
equations (\ref{RG-sep-EP}) and (\ref{RG-sep}) for the split
effective potential must be considered separately, and then we have
to extract the corresponding \textit{Type-1}\, $\mu$-dependent parts
in each of these equation. The first one contains information about
the running of the $\phi$-dependent part of the effective potential
and, therefore, about the running of the induced CC \cite{cosm},
whereas the second one includes the necessary information on the
running of an independent parameter -- the vacuum CC part:
\beq \rLV = \frac{\Lambda}{8\pi G} = hm^4\,. \label{CCd22} \eeq
Let us start the analysis with this part. The classical expressions
are related by a very simple relation
\beq U_{vac}=\rLV\,. \label{rhoCC} \eeq
So how can it be that $V_{vac}(m^2,\la,h,\mu)$ is identically
$\mu$-independent while  $\rLV$ does ``run'' with $\mu$? However, it
is exactly the situation which takes place in the MS-scheme of
renormalization. The whole point is that the relation (\ref{rhoCC})
gets modified at the quantum level. For example, at one-loop we
have, instead,
\beq  V_{vac}\,=\,\rLV(\mu) + \frac12\,
\be_\La^{(1)}\,\ln\Big(\frac{\mu_0^2}{\mu^2}\Big)\,, \label{EA-RG-1}
\eeq
where $\be_\La^{(1)}$ is the one-loop $\be$-function for the CC
term. In order to keep $V_{vac}$ identically $\mu$-independent, we
expect that the renormalized parameter $\rLV(\mu)$ ``runs'' with
$\mu$ according to
\beq \rLV(\mu) \,=\,\rLV(\mu_0) - \frac12\,
\be_\La^{(1)}\,\ln\Big(\frac{\mu_0^2}{\mu^2}\Big)\,. \label{CCdR7}
\eeq
One can easily check that these relations hold in particular cases.
For example, let us take a scalar field with mass $m$. If we compute
explicitly the corresponding vacuum-to-vacuum diagram at one-loop in
dimensional regularization, i.e. Eq.\,(\ref{rhovac}), and add up the
result to the action (\ref{dimrLO}), we easily find
\begin{equation}
\label{Vacfree}
V_{vac}=\rLV(\mu)+\delta\rLV+{\bar V}_{vac}^{(1)} \,,
\end{equation}
where the unrenormalized one-loop correction reads
\begin{equation}\label{Vacfree2}
{\bar V}_{vac}^{(1)}=\frac{m^4}{64\,\pi^2}\,\left(-
\frac{2}{4-n}-\ln\frac{4\pi\mu^2}{m^2}+\gamma_E-\frac32\right) \,.
\end{equation}
Let us adopt the $\overline{MS}$ subtraction scheme, characterized
by the counterterm
\begin{equation}\label{deltaMSB}
{\delta}\rLV=\frac{m^4}{64\,\pi^2}\,\left(\frac{2}{4-n}+\ln
4\pi-\gamma_E\right)\,.
\end{equation}
It follows that the $\overline{MS}$-renormalized vacuum part of the
potential is
\begin{equation}\label{rfvacenergy2}
V_{vac}(m,\mu)=\rLV(\mu)+{V}_{vac}^{(1)}(\mu)\,,
\end{equation}
with the finite one-loop piece
\begin{equation}\label{rfvacenergy3}
{V}_{vac}^{(1)}(\mu)=\frac{m^4}{64\,\pi^2}
\,\left(\ln\frac{m^2}{\mu^2}-\frac32\right)\,.
\end{equation}
Notice that, at one-loop, it does not depend on the self-coupling
$\lambda$. Clearly, (\ref{rfvacenergy2}) has the form
(\ref{EA-RG-1}) and we can identify
\begin{equation}\label{beta4}
\be_\La^{(1)}=\frac{m^4}{32\,\pi^2}\,.
\end{equation}
Plugging next this $\beta$-function on the \textit{r.h.s} of
Eq.\,(\ref{RGCC}), with the understanding that $\rL$ there is the
$\rLV$ part under consideration, we may easily integrate the RG
equation for the vacuum CC. {Using the fact that at one-loop the
mass $m$ does not run with $\mu$, we arrive at the result}
\begin{equation}\label{intrLV}
\rLV(\mu)=\rho^{\rm
vac}_{\La}(\mu_0)-\frac{m^4}{64\,\pi^2}\,\ln\frac{\mu_0^2}{\mu^2}\,,
\end{equation}
which confirms the general expectation (\ref{CCdR7}).

Furthermore, we can check that after replacing Eq.\,(\ref{intrLV})
into (\ref{rfvacenergy2}) we obtain a formally identical expression
in which $\mu$ has been replaced by $\mu_0$. This is the
realization, in this particular example, of the RG-invariance of the
vacuum effective potential, i.e. of Eq.\, (\ref{RG-sep}). Thus, the
$\mu$-dependence displayed on the \textit{l.h.s.} of
Eq.\,(\ref{rfvacenergy2}) is only to indicate that the various parts
on the \textit{r.h.s} of that expression are non-trivially
parameterized by $\mu$, but the sum of these parts is ultimately
independent of it. The procedure can be extended to any loop order,
and then $V_{vac}(m,\lambda)$ is in general a function of both the
mass $m$ and the self-coupling $\lambda$. \vskip 2mm

\noindent {\Large $\bullet$} \ The situation for the induced vacuum
energy density is similar. The form of the renormalized effective
potential in the $\overline{MS}$ scheme for the scalar field, with a
classical potential $U(\phi)$, is well known (see, e.g.,
\cite{Coleman85}). At one-loop, Eq.\,(\ref{Vloops}) reads (we set
$\hbar=1$ again)
\beq \EPR(\phi)=U(\phi)+ {V}^{(1)}(\phi) \,=\,U(\phi) +
\frac{1}{64\pi^2}\,{U^{\prime\prime}}^2 \,\Big(\ln
\frac{{U^{\prime\prime}}^2}{\mu^2} -\frac32\Big)\,. \label{pot} \eeq
In the case of (\ref{clpo}), the one-loop correction yields
\beq {V}^{(1)}(\phi)\,=\, \frac{1}{64\pi^2}\,\Big(m^2+\frac{\la
\phi^2}{2}\Big)^2 \,\left[\ln \frac{ \big(m^2+{\la
\phi^2}/{2}\big)}{\mu^2} -\frac32\right]\,. \label{effpo1} \eeq
To obtain $V_{scal}(\phi,m^2,\la,\mu)$, we just subtract from
(\ref{pot}) the vacuum part at one-loop, which is given by the
second term on the \textit{r.h.s.} of (\ref{rfvacenergy2}):
\begin{equation}\label{Vscalar}
{V}_{vac}^{(1)}\,=\, \frac{m^4}{64\pi^2}\,\left[\ln
\frac{m^2}{\mu^2} - \frac32\right]=\EPR(\phi=0,m^2,\la,\mu)\,,
\end{equation}
where the second equality reflects that the first expression can
also be obtained by setting $\phi=0$ in the full one-loop effective
potential (\ref{pot}). Therefore, the $\phi$-dependent part of the
potential finally reads
\begin{equation}\label{Vscal}
V_{scal}(\phi,m^2,\la,\mu)
=\EPR(\phi,m^2,\la,\mu)-\EPR(\phi=0,m^2,\la,\mu)\,.
\end{equation}
This expression satisfies Eq.\,(\ref{RG-sep-EP}). To understand why,
let us first remark the existence of an explicit (i.e.
\textit{Type-2}) $\mu$-dependence in the loop correction term
(\ref{effpo1}). For a massless case, one can restore the effective
potential from {the $\mu$-dependence} alone.

The fact that the effective potential part (\ref{Vscal}) depends
explicitly on $\mu$ and, at the same time, satisfies
(\ref{RG-sep-EP}) is because there is still the $\mu$-dependence
that is associated to the renormalization of the parameters $\la$
and $m$ (i.e. the \textit{Type-1} $\mu$-dependence). Let us indeed
recall that, once the two types of $\mu$-dependences meet together,
they annihilate each other -- see Eq.\,(\ref{Type120})\,\footnote{We
point out that, at one-loop, there is no $\mu$-dependence associated
to the field $\phi$ itself, at least for a scalar theory based on a
the classical potential (\ref{clpo}). This is a reflect of the fact
that $\gamma_{\phi}=0$ at this order in such theory.}.

For the sake of simplicity, we can illustrate the fulfillment of the
RG-invariance of $V_{scal}$ in the case when the mass $m$ is
negligible. Then, it is enough to consider the renormalization of
the $\la$ coupling, whose one-loop RG equation in the MS scheme has
the form
\beq \mu \frac{d\la}{d\mu} \, = \,\be_\la^{(1)} \, = \,
\frac{3\la^2}{(4\pi)^2}\,. \label{RGfor_la} \eeq
Integrating this equation from $\mu$ to ${\mu'}$, we obtain:
\begin{equation}\label{runmumup}
\la ({\mu'})=\frac{\la (\mu)}{1 - \displaystyle\frac{3\,\la
(\mu)}{32\,\pi^2}\, \ln\frac{{\mu'}^2}{\mu^2}}\,\simeq\, \la (\mu)+
\be_\la^{(1)}\, \tau\,,
\end{equation}
where $\tau=(1/2)\ln{{\mu'}^2/\mu^2}$. Notice that the obtained
result is similar to (\ref{renormchargeMSQEDHE2}) for the QED case.
This exemplifies, once more,  that the $\mu$-dependence of this kind
in the MS scheme is of \textit{Type-1} and, hence, it can be
{associated to} the momentum subtraction scheme in the high energy
limit.

If we replace (\ref{runmumup}) into (\ref{effpo1}) -- in the limit
of negligible $m$ -- and disregard the ${\cal O}(\la^3)$ terms that
can be affected by higher order corrections, the expression
(\ref{Vscal}) is transformed into another one which is formally
identical, but in which $\mu$ is traded for ${\mu'}$. In other
words, this confirms that $V_{scal}$ is RG-invariant to the order
under consideration, as we expected. But, again, we are not
interested in this complete RG-invariant expression for the study of
the possible sources of running. We, instead, focus only on the
implicit (\textit{Type-I}) $\mu$-dependence of $V_{scal}$, which is
obtained by replacing the parameter $m$ and $\lambda$ by the
RG-improved ones $m(\tau)$ and
$\lambda(\tau)$ at the minimum of (\ref{clpo}). The result is
\begin{equation}\label{runInd}
\rLI(\tau)=-\frac{3\,m^{4}(\tau)}{2\lambda(\tau)}\,.
\end{equation}
Let us point out that, in the context of the SM, with all the
fermions and bosons involved, the running of $\lambda=\lambda(\tau)$
is not just given by (\ref{RGfor_la}). The complete expressions for
the RG equations of the parameters $\lambda$ and $m$ within the SM
are given in Ref.\,\cite{cosm,nova}.

As explained previously, in the cosmological analysis of these
references, the correspondence of the MS with the high energy limit
just means to take into account the correct number of degrees o
freedom involved in the computation of the $\beta$-functions (for
both the vacuum and induced terms in (\ref{Lambdamu})). Namely, they
should only involve the lightest neutrinos and other light degrees
of freedom potentially present in the current Universe. In no way
presumes to consider that the present Universe is in a high energy
state! Put another way, the use of the MS scheme in the ``high
energy'' limit means, in the present context, to consider only the
contribution from particles whose masses masses satisfy $q^2> m^2$,
for $q$ of order of $\mu\sim \sqrt{H_0\,M_P}\sim 10^{-3}\,eV$. Using
(\ref{beta4}) and (\ref{runInd}), this leads, finally, to the RG
equation for (\ref{Lambdamu}) at the present cosmic scale:
\begin{equation}\label{Lambdamu2}
(4\pi)^2\,\frac{d\rL}{d\tau}
=  (4\pi)^2\,\left(\frac{d\rLV}{d\tau}+\frac{d\rLI}{d\tau}\right)
 =\frac12\,\sum_s m_s^4\,- 4\,\sum_{\nu} m_{\nu}^4 \,,
\end{equation}
where the sums extend over light scalars (e.g. the
cosmon\,\cite{PSW}) as well as over light neutrinos, all of them
satisfying the previous sharp cutoff condition. The neutrino term in
(\ref{Lambdamu2}) receives contributions from both the vacuum and
induced parts\,\cite{cosm}.

An essential observation about the cutoff $\mu\sim \sqrt{H_0\,M_P}$
is that it is determined by the value of the expansion rate at the
present time, $H_0\sim 10^{-33}\,eV$, but of course the method can
be iterated to values of $H$ at any time in the history of the
Universe, see the analysis of \cite{nova}. This fact clearly shows
that, in the absence of an expanding background, the whole procedure
ceases to make sense. This crucial aspect went also completely
unnoticed by the authors of Ref.\,\cite{AG}.
Despite its limitations, the H-dependence of the
cutoff is an effective tool that the MS-scheme based RG possesses to
explore the possible dependence of the EA on the external field. In
more formal terms: the presence of the external metric is the actual
source for the possible running. The relevant Feynman diagrams
include matter fields loops with a number of external tails of the
metric (see, e.g., \cite{PoImpo}). In this situation, even if we can
only rely upon the MS-renormalization scheme, there are many choices
for a physically reasonable identification of $\mu$. We used some
possibilities in \cite{cosm,nova} -- one of them has just been
mentioned above. However, at variance with the flat space-time case,
there is not a single obvious choice comparable to the S-matrix
problem in high energy physics, where $\mu$ can be taken of order of
the typical energy of the scattered particle.

\vspace{0.2cm}

\noindent {\Large $\bullet$}  The observed CC should emerge from the
complete (and RG-invariant) theoretical expression (\ref{CCi}),
provided it would be really feasible to explicitly account for it
(i.e. if it would be possible to solve the ``old CC problem'' in
some concrete TOE framework! -- see section 1). However,
accomplishing this final aim, is well out of the scope of the RG
analysis, just because of the RG-invariance of the CC as a physical
observable! Let us recall the reader, once more, that the RG
considerations are framed in the context of the \textit{BII} kind of
approach to the CC problem. Therefore, the partial sum
(\ref{Lambdamu}) should suffice to fulfil the (much more modest) RG
aim since it already carries the real ``signature'' of the possible
running, namely it includes the \textit{Type-I} $\mu$-dependence.
Why only this $\mu$-dependence? Because {\it the difference with the
complete expression does not matter for the running!} The dependence
of the EA on the metric and its derivatives does not change if we
sum up the two types of contributions. However, by summing them, we
lose the possibility to employ the RG (that is, the
$\mu$-dependence) to unveil the quantum structure of the EA.
\vspace{0.2cm}

We end up this section with a few additional comments, which are
also relevant:

\noindent {\Large $\bullet$} If the considerations of \cite{AG}
would be correct, it would mean, in particular, that we could define
both induced constants CC and $G$ in a unique way. If so, this could
solve a long-standing problem of an ambiguity in the action of
induced gravity \cite{Adler}. Of course, the real situation is just
the opposite, that is the ``proof'' given in \cite{AG} is empty of
content and the ambiguity is there. The latter is associated to the
dynamical breaking of scale invariance, hence to the trace anomaly.
The induced vacuum energy reads, in the general case,
\begin{equation}\label{TraceA}
\rLI(\mu)=\frac{\La^{ind}(\mu)}{8\pi\,G^{ind}(\mu)}=\frac14\,\langle
T_{\alpha}^{\alpha}(\mu)\rangle\,,
\end{equation}
where the trace on the \textit{r.h.s.} contains both the classical
and quantum effects, which may allow for the running. Let us
remember that the mentioned ambiguity means that both CC and $G$
depend on a single parameter which can not be defined in the
framework of the initial quantum theory of matter fields. This
dependence, according to \cite{Adler} is universal, it has a deep
physical sense and means that the induced CC and $G$ can, in
principe, run with the cosmic scale.

\vspace{0.2cm}

\noindent {\Large $\bullet$} We have already seen above, that the
presence of a non-trivial metric can essentially change the
situation, in contrast to the claims of \cite{AG}. But there is more
to say. For example, in a curved space-time, the minimal interacting
scalar field is not renormalizable and hence the MS-based
renormalization group can not be applied \cite{book}. If, however,
we admit the presence of a non-minimal interaction of the scalar
field with the curvature, the spontaneous symmetry breaking (SSB)
becomes rather nontrivial and the renormalization of the theory is
very complicated \cite{sponta}. Moreover, the SSB generates an
infinite number of non-local terms in the induced action of gravity.
So, what looks somehow trivial in the simpleminded framework of
\cite{AG}, in reality is not so simple.

\vspace{0.2cm}

After making transparent that the ``proof'' of the non-running CC,
as presented in \cite{AG}, is insubstantial (in fact, false) in all
its parts, we still have to answer the following most important
question:  do all the arguments presented above mean that the
induced CC and $G$, or their vacuum counterparts, really do run with
the cosmic scale? The answer is negative. Let us repeat that the
renormalization group running is nothing else but a clever way to
parameterize the quantum corrections. Therefore, in order to really
establish the running we have to derive these corrections, or at
least indicate the form which they can have. While in the  MS-based
RG model of \cite{cosm}, the non-trivial properties of the FLRW
metric are encoded in the $H$-dependent cutoff $\mu\sim
\sqrt{H\,M_P}$, one would like to have a more direct (and reliable)
approach to the alleged running properties of the physical CC. In
other words, one would like to have an expression for (\ref{CCi}) in
which the $H$-dependence is explicit, and where the decoupling
effects would be manifest. A phenomenological model describing this
possibility is represented e.g. by Eq.\,(\ref{expecVERen}). However,
as we have said, at the moment there is no fundamental derivation of
this formula.  In the next section, we address some possible roads
leading to this kind of more physical relations, after studying and
classifying the possible forms of the quantum effects on the EA in
the context of the momentum subtraction scheme.

\section{Possible forms of the quantum effects: roads to the
physical running of the cosmological parameters}

\qquad The purpose of this section is to discuss the possible form
of the quantum corrections to the Einstein-Hilbert action and to the
CC term. As we have already discussed above, such corrections should
be represented by some functional of the metric and its derivatives,
while other fields are in the stable vacuum states. Therefore, in
this case we do not need to distinguish the vacuum and induced parts
of the quantum corrections.

The classical action of vacuum for quantum matter fields is
contained in the full action (\ref{total}) as follows: \beq S_{vac}
&=& S_{EH}\,+\,S_{HD}\,, \label{vacuum} \eeq where $\,S_{EH}\,$ is
the Einstein-Hilbert action with the cosmological constant and the
second term represents a minimal set of higher derivatives necessary
to insure the consistency of the theory at the quantum level: \beq
S_{HD} &=& \int d^4x \sqrt{-g} \left\{a_1C^2+a_2E+a_3{\Box}R+a_4R^2
\right\}\,. \label{HD} \eeq Here $\,C^2=R_{\mu\nu\al\be}^2 - 2
R_{\al\be}^2 + (1/3)\,R^2\,$ is the square of the Weyl tensor and
$\,E = R_{\mu\nu\al\be}^2 - 4 R_{\al\be}^2 + R^2\,$ is the integrand
of the Gauss-Bonnet topological term.

Looking at the expressions (\ref{total}),(\ref{vacuum})  and
(\ref{HD}), one can ask a most natural question: how should we
distribute the quantum corrections -- which are typically given by
non-local and non-polynomial expressions the in curvature tensor --
between the different terms in the overall EA of vacuum? The first
option is to perform explicit calculations of the quantum
corrections, including the finite part. The history of such
calculations goes back to the early works on quantum theory in
curved space \cite{stze71}. The most complete result in this
direction has been obtained in \cite{apco} through derivation of
Feynman diagrams in the framework of linearized gravity and also by
using the heat kernel solution of \cite{bavi90}. The output of such
calculation includes only the terms of zero, first and second order
in the curvature tensor.

At this level, one can identify the form factors of different terms
of the quantum analog of the action $S_{vac}$ in (\ref{vacuum}) by
means of analyzing their tensor structure. For example, the one-loop
result for the massive real scalar field has the form \cite{apco}
\beq
{\bar \Ga}^{(1)}_{vac}
&=&
\frac{1}{2(4\pi)^2}\,\int d^4x \,\sqrt{-g}\,
\Big\{\,\frac{m^4}{2}\cdot\Big(\frac{1}{\ep} +\frac32\Big)
\,+ \,\Big(\xi-\frac16\Big)\,m^2R\, \Big(\frac{1}{\ep}+1\Big)
\nonumber
\\
&+&
\frac12\,C_{\mu\nu\al\be} \,\Big[\frac{1}{60\,\ep}+k_W(a)\Big]
C^{\mu\nu\al\be} \,+\,R
\,\Big[\,\frac{1}{2\ep}\,\Big(\xi-\frac16\Big)^2\, +
k_R(a)\,\Big]\,R\,\Big\}\,.
\label{final}
\eeq
Here $\,\ep\,$ is the parameter of dimensional regularization,
\beq \frac{1}{\ep}=\frac{2}{4-n} +\ln \Big(\frac{4\pi
\mu^2}{m^2}\Big) - \ga_E\,, \label{epsilon} \eeq
and $\xi$ is the coefficient of the non-minimal coupling. From
(\ref{VeffEA1}) it is clear that the first term on the
\textit{r.h.s.} of (\ref{final}) coincides with $-{\bar
V}_{vac}^{(1)}$, see Eq.\,(\ref{Vacfree2}), i.e. the unrenormalized
one-loop vacuum correction to the effective potential case
considered in section 4, whose renormalized form in the
$\overline{MS}$ scheme is the expression ${V}_{vac}^{(1)}$ given by
(\ref{rfvacenergy3}). The full result (\ref{final}), therefore,
provides the corresponding generalization for the case when there is
a non-trivial background. The presence of the curvature terms and
the non-local form factors $k_W(a)$ and $k_R(a)$ -- see their
structure below -- clearly indicates that one obtains a highly
non-trivial generalization of the simple flat space-time expression
studied in section 4. The essential observation is that it is no
longer possible to perform an effective potential approach because
some of the new terms provide non-local contributions which are
explicitly dependent on the external momenta.

The mentioned form factors read, explicitly,
\beq k_W(a) &=& \frac{8A}{15\,a^4}
\,+\,\frac{2}{45\,a^2}\,+\,\frac{1}{150}\,, \nonumber
\\
k_R(a) &=&
A\Big(\xi-\frac16\Big)^2-\frac{A}{6}\,\Big(\xi-\frac16\Big)
+\frac{2A}{3a^2}\,\Big(\xi-\frac16\Big)
+\frac{A}{9a^4}-\frac{A}{18a^2}+\frac{A}{144}+
\nonumber
\\
&+& \frac{1}{108\,a^2} -\frac{7}{2160} +
\frac{1}{18}\,\Big(\xi-\frac16\Big)\,, \label{W} \eeq
where we used
notations
\beq A\,=\,1-\frac{1}{a}\ln\,\Big(\frac{2+a}{2-a}\Big)\,, \qquad a^2
= \frac{4\Box}{\Box - 4m^2}\,. \label{Aa} \eeq
Similar expressions have been obtained for massive fermion and
vector cases \cite{apco} and also for the scalar field background,
where the form factors were calculated for the $\,\phi^2 R\,$ and
$\,\phi^4\,$ terms \cite{bexi}. In the UV limit all form factors
have, in general, the logarithmic behavior similar to (\ref{QED}),
but in the IR limit they follow the Appelquist and Carazzone theorem
\cite{apco,PoImpo}.

The expressions (\ref{W}) suggest how
the desirable quantum corrections responsible for the physical
running could, in principle, look like: \ they should form a
non-local and complicated expression, but still a relatively compact
one. The form factors (\ref{W}) contain a lot of information about
quantum corrections in the vacuum sector, including the low-energy
decoupling, conformal anomaly in the massless limit of the theory
and so on \cite{PoImpo}. However, when looking at the CC and
Einstein-Hilbert sectors of (\ref{final}), our optimism somehow
stagnates because there is no relevant form factor in these
sectors. In both cases there are divergences, there is the
$\mu$-dependence (as it should be, of course), but there is
nothing real behind this $\mu$-dependence.
Indeed, we have mentioned above that the CC part is
identical to the MS-based renormalization group analysis that we
have presented in section 4, namely the $\mu$-dependence of these
terms is just of \textit{Type-2}, see section 2.3. As a result, they
have no correspondence with the physical renormalization group in
the UV limit. In contrast, in the higher derivative sectors, we meet
a perfect correspondence with the momentum subtraction scheme.
Notice that, in momentum space, the form factors (\ref{W}) are
functions of momentum $q$ through (\ref{Aa}), where we have the
correspondence $a^2\to 4\,q^2/(q^2+4\,m^2)$.

As compared to the higher derivative terms, the origin for the
unpaired result in the CC and EH case is that, when the the CC is
present, the expansion around the flat background is not a perfect
instrument for obtaining the quantum corrections. The form factors
of the higher derivative terms do not change when we pass to the
flat space background, but the ones for the CC and Einstein-Hilbert
terms just vanish, because the form factors should be constructed
from the d'Alembert operators, while $\Box\La=0$ and $\Box R$ is an
irrelevant total derivative term. This is an explicit manifestation
of the unfortunate property of the calculation based on the
expansion around the flat-space. This calculational scheme can not
provide information about non-local corrections to the CC and EH
terms, independent of whether such corrections really exist. It is
important to emphasize that there is no better technique available
and, moreover, it is rather unclear how such technique may look. It
is at least clear that such technique can not be related to the
expansion in the powers in curvatures, because such expansion is not
going to be efficient beyond the flat-space expansion framework.
However, the lack of the appropriate technique by no means implies
that there is no physical effect. Let us invoke, once again, the
example of QCD. We know that the perturbative expansion is not
efficient at the energies below, e.g. $1\,MeV$, but this does not
mean that the low-energy QCD effects are irrelevant!

A good hint that the renormalization
group corrections to the CC and
Einstein-Hilbert terms are possible, can be obtained within the
method suggested in \cite{Shocom} (see also
\cite{asta}). The idea is to perform a conformization
procedure\,\cite{PSW} and derive the anomaly induced EA. We refer to
the mentioned papers for details and just present the final result
at the one-loop level
$$
{\bar \Gamma} \,\,=\,\,S_c[g_{\mu\nu}] \,-\, \frac{3c+2b}{36}\,\int
d^4x\sqrt{-g}\,[{\bar R} - 6({\bar \nabla}\sigma)^2 - 6({\bar
{\nabla}}^2 \sigma)]^2
$$
\vskip 1mm
$$
\,+\, \int d^4 x\sqrt{-{\bar g}} \,\{w{\bar C}^2\sigma \,+\,
b\,({\bar E} -\frac23 {\bar \nabla}^2 {\bar R})\,\sigma \,+\, 2
b\,\sigma{\bar \Delta_4}\sigma \}
$$
\beq
-\,\int d^4 x\sqrt{-{\bar g}} \,e^{2\si}\, [{\bar R}+6({\bar
\na}\si)^2] \,\cdot\, \Big[\, \frac{1}{16\pi G} - f\cdot\si\,\Big] -
\int d^4 x\sqrt{-{\bar g}}\,e^{4\si}\,\cdot\, \Big[\frac{\La}{8\pi
G}\,-\,g\cdot\si\,\Big] \,,
\label{quantum for massive}
\eeq
where
$S_c$ is an unknown integration constant for the EA, and
\begin{eqnarray}
\label{Delta4}
\Delta_4 = \nabla^4 + 2\,R^{\mu\nu}\nabla_\mu\nabla_\nu -
\frac23\,R{\nabla^2} + \frac13\,(\nabla^\mu R)\nabla_\mu
\end{eqnarray}
is the fourth order, self-adjoint, conformal operator acting on
scalars.

The action (\ref{quantum for massive}) is written using a special
conformal parametrization of the metric $g_{\mu\nu}=e^{2\si(x)}{\bar
g}_{\mu\nu}$, similar to the Eq.\, (\ref{conf_rep}). Here
$\om,\,b,\,c,\,g,\,f$ are the MS-based $\be$-functions for the
parameters of the vacuum action (\ref{vacuum}).
For example, $f$ and $g$ are dimensional quantities associated
to the renormalization of the CC and EH terms:
\beq
f\,=\,\frac{1}{3(4\pi)^2}\,\sum_{f}\,{N_f\,m_f^2}\,, \label{f}
\eeq and \beq g\,=\,\frac{1}{2(4\pi)^2}\,\sum_{s} \,{N_s\,m_s^4}
-\frac{2}{(4\pi)^2}\sum_{f}\,{N_f\,m_f^4}\,.
\label{g}
\eeq
Here the sums are taken over all massive fermion $f$ and scalar $s$
fields with the masses $m_f$ and $m_s$ correspondingly; $N_f$ and
$N_s$ are multiplicities of fermions and scalars.

As noticed in \,\cite{Shocom}, the above expression (\ref{quantum
for massive}) is exactly the local generalization of the
renormalization group corrected classical action (\ref{vacuum}).
The last is
defined through the solution of the RG equation for the effective
action, i.e. the solution of the PDE (\ref{RGGamma}). Such solution
is well-known to follow from the method of characteristics and is
expressed in terms of the running charges $P(\tau)$ and fields
${\Phi}(\tau)$, see section 2.3. Using a notation similar to
(\ref{nn8}), the RG transformation on the renormalized EA in the MS
scheme leads to
\begin{eqnarray}
\Gamma[e^{-2\tau}g_{\alpha\beta},{\Phi}(\mu),P(\mu),\mu, n ] =
\Gamma[g_{\alpha\beta},{\Phi}(\tau),P(\tau),\mu, n ]\,, \label{RGEA}
\end{eqnarray}
where, as before, $\tau=(1/2)\,\ln ({\mu'}^2/\mu^2)$. In the
leading-log approximation, one can take, instead of (\ref{RGEA}),
the classical action and replace (for the massless conformal theory)
$P(\mu)$ with $P(\mu')= P(\mu) + \beta_P\,\tau$, where $\beta_P$ is
the $\beta$-function for the parameter $P$. A particular example of
this expression for $P=\lambda$ is Eq.\,(\ref{runmumup}), with the
$\beta$-function (\ref{RGfor_la}). Upon comparing (\ref{quantum for
massive}) with the result of this procedure, one confirms the
complete equivalence of the two expressions in the terms that do not
vanish for $\sigma=const$. The important general conclusion is that
the procedure \cite{Shocom,asta} leading to the effective action
(\ref{quantum for massive}) is consistent with the anomaly-induced
effective action and is indeed a direct generalization of the
RG-improved classical action.

Let us point out that the expression (\ref{quantum for massive}) has
been obtained in the framework of the minimal subtraction
renormalization scheme, which is supposed to converge to the
physical renormalization scheme at high energies. Therefore, the
absence of the relevant form factors in the low-energy sectors of
the expression (\ref{final}) looks as an indication that the method
used for deriving this expression (being equivalent to the expansion
near the flat space metric) is not an appropriate instrument for
deriving the quantum correction to these terms.  \ In contrast, the
derivation of (\ref{quantum for massive}) does not directly rely on
the expansion near the flat space metric and hence should be
regarded as a kind of indirect confirmation of the existence of a
nontrivial quantum corrections to CC and Einstein-Hilber terms.

In the situation when the perspective to obtain a solid quantum
field theory based result is unclear, it is absolutely legitimate
to use the phenomenological approach \cite{cosm}-\cite{fossil},
investigating the role of such possible running and trying to
impose some cosmological restrictions on it. However, a simple
albeit
very important criticism of the CC running has been presented
by Dolgov \cite{Dolgov}, but in fact goes back to the epoch of
L.D. Landau. If we consider that the energy-momentum tensor
of the vacuum is given by (\ref{EMT}) and (\ref{CCd2}), i.e.
\beq (T_{\CC})^\al_\be =
\mbox{diag}\,(\rho_\La,\,\rho_\La,\,\rho_\La,\,\rho_\La) \,, \eeq
the conservation equation immediately tells us that
$\rho_\La=const$, as we have discussed in section 2.2.  However, if
we attribute a scale dependence to the CC and realize that the
cosmic energy scale changes with time, we arrive at the
time-dependent CC \cite{nova,CCfit,Gruni}, apparently contradicting
the condition $\rho_\La=const$. There are nevertheless some
significant loopholes in this argument. In section 2.2, we have
already seen that the conservation law can be modified such that the
CC is time dependent -- through e.g. some relation of the type
(\ref{dLdt}) -- at the price of admitting non-conservation of matter
or even variable gravitational coupling $G$. But, most important,
even if there is no direct time-dependence of these quantities, the
possibility of scale-dependence, as indicated in Eq.\,(\ref{RGCC}),
can also play a crucial role.

Indeed, let us remark that, in the case of quantum corrections in
curved space-time, the aforementioned conservation equation just
reflects the covariance of the effective action. However, the
covariance of nontrivial quantum corrections to the CC typically
takes the form of nonlocal terms (see below). One can construct an
infinite amount of non-local covariant actions and there is no
guarantee that some of them will not give the same effect as a
running CC and/or as a running $G$. As a particular example, we have
to mention the anomaly-induced EA for the massless case, that is,
the first two lines of  Eq.\,(\ref{quantum for massive}), which give
rise to the Starobinsky model of inflation \cite{star} even without
the CC term!

The covariance is the fundamental guiding principle in the study of
vacuum quantum corrections \cite{birdav,book}. One of the important
consequences of the covariance of the EA is that, when performing
the expansion of this functional in the metric derivatives, there
can not be terms which are odd in these derivatives
\cite{CCfit,PoImpo}. We have seen an example in the ansatz
(\ref{expecVERen}). In the cosmological setting, this implies that
the quantum correction to the CC term has to be constructed from the
curvature tensor, its covariant derivatives and also from the
Green's functions of some covariant operators. At the low-energy
cosmological scale, we can use the Hubble parameter $H$ as the
measure of this scale \cite{nova,CCfit,Gruni,fossil}; in this case,
the corrections can start from terms of ${\cal O}(H^2)$, but not
from ${\cal O}(H)$.

The last feature suggests that the correction to the CC term in the
current Universe has the form (\ref{expecVERen}), where the dominant
effect is $\De \rho_\La \sim s\,M^2\,H^2$, with $M^2$ an average sum
of the contributions of all massive particles which are present in
both the observable and unobservable parts of the particle spectrum
(e.g. including GUT fields with very large masses near the Planck
mass); and $s=\pm 1$, depending on whether bosons or fermions
dominate in the higher end of such spectrum. This form of quantum
corrections correspond to the Appelquist and Carazzone theorem
\cite{nova,babic}.

At this point we meet the following natural question, formulated by
A. Vilenkin \cite{Vilenkin}: \
{\it In view of the fact that  $R\sim H^2$, the
dominant correction would be proportional to the curvature scalar,
$\De \rho_\La\sim M^2\,R$, and one could naively conclude that it is
not a non-trivial correction to the CC term, but rather an additive
contribution to the Einstein-Hilbert term.} Again, the solution of
this puzzle can be possible only because the relevant quantum
effects must include nonlocalities (see below) and, thus, in general
we  expect that it cannot reproduce any of the local sectors of the
classical action $S_{vac}$ (\ref{vacuum}).

An example of such non-local contribution is the term in the
expression (\ref{quantum for massive}) containing the  fourth order
conformal operator $\De_4$. Due to the presence of the fiducial
metric $\bar{g}_{\mu\nu}$, rather than the original one
${g}_{\mu\nu}$, the expression (\ref{quantum for massive}) although
it looks relatively simple and compact is not manifestly covariant.
Interestingly enough, when written in a covariant form (see, e.g.
\cite{rie,PoImpo}), the anomaly induced action includes the Green's
functions insertions of the operator $\De_4$ and hence one meets a
manifestly non-local expression.  As an example, the term
proportional to the Weyl tensor squared, \ $-\,\bar{C}^2$, \
in (\ref{quantum for massive}) can be written in a covariant,
but non-local, form as follows:
\beq
\frac{1}{4}\,\int d^4 x \sqrt{-g
(x)}\, \int d^4 x' \sqrt{-g (x')}\, C^2(x) \,G(x,x')\,(E -
\frac23{\Box}R)_{x'}\,, \label{xxp}
\eeq
where $G(x,x')$ is the Green function of $\De_4$, i.e.
$\De_{4,x}\,G(x,x')=\de(x,x')$.

The \ $4d$ \ expressions which are present in the vacuum EA, are
similar to the well known Polyakov action in \ $2d$ \cite{odsh91}.
However, in the present case there are very strong additional
restrictions, concerning the quantum corrections to the CC and
Hilbert-Einstein term. For example, these quantum contributions can
not emerge in terms which are quadratic in curvatures. In order to
prove this statement, let us note that the terms which do not
reproduce the form of the classical vacuum action (\ref{vacuum}),
should be at least ${\cal O}(R^2_{....})$ and also have some Green's
functions insertions. These insertions should correspond to the
massive quantum field propagator in curved space, but since the
${\cal O}(R^2_{....})$  terms admit the flat space expansion,
$g_{\mu\nu}=\eta_{\mu\nu}+h_{\mu\nu}$, we can take the ${\cal
O}(h_{\mu\nu})$ term for each curvature and flat-space propagator.
Hence, to the order ${\cal O}(R^2_{....})$, the result is given by
the second line of Eq. (\ref{final}) and its low-energy limit is
characterized by the Appelquist and Carazzone - like decoupling (see
detailed discussion in \cite{apco,PoImpo}). In order to understand
the phenomenological consequences, let us consider two simplified
expressions which possess, qualitatively, the same property \beq
R_{\mu\nu}\,\frac{1}{\Box + m^2}\,R^{\mu\nu} \,\,, \qquad
R\,\frac{1}{\Box + m^2}\,R\,. \label{nonloc} \eeq As far as we are
always interested in the large mass limit, we can expand the
propagator as \beq \frac{1}{\Box + m^2} \sim \frac{1}{m^2}\, \Big(1
- \frac{\Box}{m^2} + \,...\Big)\,, \label{propa} \eeq The first term
in the parenthesis obviously gives just an additive contribution to
the local structures in the vacuum action (\ref{vacuum}).
Furthermore, the second term leads to the ${\cal O}(H^6)$
contribution and is, thus, phenomenologically irrelevant.

Let us go to higher orders in the curvature expansion. It is clear
that the terms starting from ${\cal O}(R^3_{....})$, in the
low-energy cosmological setting are at least ${\cal O}(H^6)$ in the
case where there are only the  insertions of massive Green's
function. Similarly, \ ${\cal O}(R^4_{....})\sim {\cal O}(H^8)$ etc.
In order to obtain the relevant running of the CC and $1/G$ from the
quantum corrections, a resummation in these terms must occur such
that the massive Green's functions are traded for the massless ones.
Therefore, we expect that the resummed result should generate a
non-local contribution to the CC of the form $R\, {\cal F}(G_0\,
R)$, for some unknown function ${\cal F}$ of dimension $2$, where
$G_0$ is the massless Green's function ($G_0\sim
1/\Box$)\,\cite{DeserWood07}. The canonical possibility would be
${\cal F}= M^2 G_0 R$, where $M$ is the mass of the particular field
involved in this term. Each one of these non-local contributions
should be added to the \textit{r.h.s.} of Eq.\,(\ref{CCi}).
Obviously, the heaviest fields would be dominant in that kind of
non-local quantum effects, so $M$ should mainly stand for the GUT
masses of order of the Planck mass $M_P$. It is easy to see that
this would provide an effective running CC behavior of the form
$\sim M_P^2 H^2$.

It is important that such resummation, in principle, may take place
in any higher order of curvature expansion or in the sum of the
infinite power series of the curvature expansion. Let us note that a
similar resummation does in fact take place in the anomaly-induced
EA of vacuum for the massless conformal case \cite{Deser93}. Due to
such resummation, the exact compact expression for this EA (see
\cite{Conf-Proc} for detailed discussion) does not involve the
Green's functions of the original quantum fields (scalars, spinors
and vectors) but only the Green's function of the higher derivative
conformal operator (\ref{Delta4}). Also, the power of the curvature
expansion of the relevant terms changes from third to fourth. Let us
note that this occurrence looked impossible at the early stage of
investigating the form of the anomaly-induced EA \cite{DesSch}. 

It has been known for a long time \cite{stze71} that the situation
for massive fields must be essentially more complicated and
cumbersome compared to the massless case. It might happen that the
form of the quantum contribution of our interest can not be derived
in a covariant form, especially because the practical use of
covariant methods implies some finite order in curvature expansion.
Therefore, in view of the difficulty of practical deriving the
higher order corrections, the program to prove or disprove the
existence of the relevant running of the CC and $1/G$ does not look
{straightforward}. In fact, the complicated form of the possible
quantum corrections to the CC and Einstein-Hilbert terms just
indicates to us (once again) that the expansion into series in
curvatures is not a proper instrument for exploring the possible
corrections to these terms.

Finally, from the above considerations it follows that the unique
conclusion we may draw from covariance is that, if the relevant
quantum corrections to the CC do exist, they have the following
effective form (after imposing the boundary condition at the present
time)
\begin{equation}
\label{runphysCC}
\rL^{ph}(H)= \rL^{0}+ \beta\,M_P^2\,(H^2-H_0^2)+ {\cal O}(H^4)\,,
\end{equation}
where $\rL^{0}$ is the current value of the CC and $\beta$ some
numerical coefficient. This expression is precisely of the
``soft-decoupling'' form (\ref{expecVERen}).

To summarize, the present day theoretical methods do not enable us
to make a conclusive verdict about the running, except the universal
form of it (\ref{runphysCC}). The problem is of course very
important and future work in this area should combine both things:
improving our theoretical methods and also continue to work on a
better understanding of the phenomenological models based on the
possible CC running \cite{cosm}-\cite{fossil}.

\section{Conclusions}

\qquad We have considered various quantum field theory aspects of
the running of the CC term or vacuum energy density. The existing
field theoretical methods can not prove that such running takes
place but cannot disprove such running either. While we have no
definite formal conclusion concerning the CC running, all positive
statements about its existence must be done with a proper caution.
However, the same is true for the negative statements, because
trying to prove ``non-running'' theorems within an incorrect
approach leads to severe errors. In particular, one such errors was
committed in the recent paper \cite{AG}, where a non-running
statement was incorrectly supported by (extremely naive) arguments
based on the $\mu$-independence of the EA in the framework of the
MS-based renormalization scheme. In reality, such $\mu$-invariance
of EA is an automatic property which holds for all kinds of
renormalizable theories and, definitely, can not be used to derive
any conclusions on the running.

The possibility of having a physically measurable CC running and, in
general, of measuring the low-energy quantum corrections coming from
massive fields is, in our opinion, one of the most important and
most interesting applications of quantum field theory in the years
to come. The resolution of the corresponding theoretical problems is
difficult, but does not look completely inaccessible. The
identification of the possible form of the vacuum effects of quantum
massive fields is the key issue for the understanding of whether the
variable DE in our Universe can be the consequence of the known
fundamental physics or only the manifestation of a qualitatively new
entities such as quintessence, k-essence, Chaplygin gas etc. Should
the analysis of the astronomical data eventually give preference to
a mildly variable DE, then this problem would become one of the most
important ones of our time. In such case, the possibility that the
cosmological parameters are running quantities would constitute the
most relevant interface between quantum field theory and
cosmology\,\cite{nova}.

\section*{Acknowledgements}
The authors are thankful to A. Barvinsky, A. Dolgov, E. Gorbar,
B. Guberina, R. Horvat, A. Starobinsky, H. Stefancic, A. Vilenkin
and E. Verdaguer
for helpful discussions. The work of I.S. has been supported by CNPq
(Brazil), FAPEMIG (MG, Brazil), FAPESP (ES, Brazil) and ICTP. The
work of J.S. has been supported in part by MEC and FEDER under
project FPA2007-66665 and also by DURSI Generalitat de Catalunya
under project 2005SGR00564 and  by the Spanish Consolider-Ingenio
2010 program CPAN CSD2007-00042. J.S. would also like to thank the
Max Planck Institut f\"ur Physik in Munich for the hospitality and
financial support while part of this work was being carried out, and
also to the Brazilian agency FAPEMIG for the support provided during
a visit at the Dept. of Physics of the Univ. Federal de Juiz de
Fora.

\newcommand{\JHEP}[3]{ {JHEP} {#1} (#2)  {#3}}
\newcommand{\NPB}[3]{{\sl Nucl. Phys. } {\bf B#1} (#2)  {#3}}
\newcommand{\NPPS}[3]{{\sl Nucl. Phys. Proc. Supp. } {\bf #1} (#2)  {#3}}
\newcommand{\PRD}[3]{{\sl Phys. Rev. } {\bf D#1} (#2)   {#3}}
\newcommand{\PLB}[3]{{\sl Phys. Lett. } {\bf B#1} (#2)  {#3}}
\newcommand{\EPJ}[3]{{\sl Eur. Phys. J } {\bf C#1} (#2)  {#3}}
\newcommand{\PR}[3]{{\sl Phys. Rep } {\bf #1} (#2)  {#3}}
\newcommand{\RMP}[3]{{\sl Rev. Mod. Phys. } {\bf #1} (#2)  {#3}}
\newcommand{\IJMP}[3]{{\sl Int. J. of Mod. Phys. } {\bf #1} (#2)  {#3}}
\newcommand{\PRL}[3]{{\sl Phys. Rev. Lett. } {\bf #1} (#2) {#3}}
\newcommand{\ZFP}[3]{{\sl Zeitsch. f. Physik } {\bf C#1} (#2)  {#3}}
\newcommand{\MPLA}[3]{{\sl Mod. Phys. Lett. } {\bf A#1} (#2) {#3}}
\newcommand{\JPA}[3]{{\sl J. Phys.} {\bf A#1} (#2) {#3}}
\newcommand{\CQG}[3]{{\sl Class. Quant. Grav. } {\bf #1} (#2) {#3}}
\newcommand{\JCAP}[3]{{ JCAP} {\bf#1} (#2)  {#3}}
\newcommand{\APJ}[3]{{\sl Astrophys. J. } {\bf #1} (#2)  {#3}}
\newcommand{\AMJ}[3]{{\sl Astronom. J. } {\bf #1} (#2)  {#3}}
\newcommand{\APP}[3]{{\sl Astropart. Phys. } {\bf #1} (#2)  {#3}}
\newcommand{\AAP}[3]{{\sl Astron. Astrophys. } {\bf #1} (#2)  {#3}}
\newcommand{\MNRAS}[3]{{\sl Mon. Not. Roy. Astron. Soc.} {\bf #1} (#2)  {#3}}


\end{document}